\documentclass[prb,longbibliography,twocolumn]{revtex4-1}

\usepackage[utf8]{inputenc}
\usepackage{amssymb}
\usepackage{amsmath}
\usepackage{amsfonts}
\usepackage[usenames,dvipsnames]{xcolor}
\usepackage[pdftex]{graphicx}
\usepackage{bm}
\usepackage{hyperref}
\newenvironment{mpmatrix}{\begin{medsize}\begin{pmatrix}}%
{\end{pmatrix}\end{medsize}} % Use accordingly to nccmath 
\usepackage{nccmath} % To resize the matrices (creating new environments).

\hypersetup{pdfa,plainpages=false,colorlinks=true,linkcolor=Red, citecolor=blue, urlcolor=blue}

\usepackage{array}
\newcolumntype{L}[1]{>{\raggedright\let\newline\\\arraybackslash\hspace{0pt}}m{#1}}
\newcolumntype{C}[1]{>{\centering\let\newline\\\arraybackslash\hspace{0pt}}m{#1}}
\newcolumntype{R}[1]{>{\raggedleft\let\newline\\\arraybackslash\hspace{0pt}}m{#1}}

\newcommand{\angstrom}{\mbox{\normalfont\AA}}
\newcommand*{\Scale}[2][4]{\scalebox{#1}{$#2$}}

\begin{document}
% Title layout
\title{Superfluid stiffness in cuprates: Effect of Mott transition and phase competition}
\author{O. \surname{Simard}}
%\email[Corresponding author: ]{simon.verret@usherbrooke.ca}
\affiliation{D\'epartement de physique and Institut quantique, Universit\'e de Sherbrooke, Qu\'ebec, Canada  J1K 2R1}
\author{C.-D. \surname{Hébert}}
\affiliation{D\'epartement de physique and Institut quantique, Universit\'e de Sherbrooke, Qu\'ebec, Canada  J1K 2R1}
\author{A. \surname{Foley}}
\affiliation{D\'epartement de physique and Institut quantique, Universit\'e de Sherbrooke, Qu\'ebec, Canada  J1K 2R1}
\author{D. \surname{S\'en\'echal}}
\affiliation{D\'epartement de physique and Institut quantique, Universit\'e de Sherbrooke, Qu\'ebec, Canada  J1K 2R1}
\author{A.-M. S. \surname{Tremblay}}
\affiliation{D\'epartement de physique and Institut quantique, Universit\'e de Sherbrooke, Qu\'ebec, Canada  J1K 2R1}
\affiliation{Canadian Institute for Advanced Research, Toronto, Ontario, Canada M5G 1Z8}
\date{\today}
%  Abstract
\begin{abstract}
Superfluid stiffness $\rho_s$ is a defining characteristic of the superconducting state, allowing phase coherence and supercurrent. It is accessible experimentally through the penetration depth. Coexistence of $d$-wave superconductivity with other phases in underdoped cuprates, such as antiferromagnetism (AF) or charge-density waves (CDW), may drastically alter $\rho_s$. To shed light on this physics, the zero-temperature value of $\rho_s=\rho_{zz}$ along the $c$-axis was computed for different values of Hubbard interaction $U$ and different sets of tight-binding parameters describing the high-temperature superconductors YBCO and NCCO. We used Cellular Dynamical Mean-Field Theory for the one-band Hubbard model with exact diagonalization as impurity solver and state-of-the-art bath parametrization. We conclude that Mott physics plays a dominant role in determining the superfluid stiffness on the hole-doped side of the phase diagram. On the electron-doped side, antiferromagnetism wins over superconductivity near half-filling. But upon approaching optimal electron-doping, homogeneous coexistence between superconductivity and antiferromagnetism causes the superfluid stiffness to drop sharply. Hence, on the electron-doped side, it is competition between antiferromagnetism and $d$-wave superconductivity that plays a dominant role in determining the value of $\rho_{zz}$ near half-filling. At large overdoping, $\rho_{zz}$ behaves in a more BCS-like manner in both the electron- and hole-doped cases. We comment on some qualitative implications of these results for the superconducting transition temperature. 

% When $\rho_s$ is small, phase coherence may occur at a lower temperature than Cooper pair formation, lowering the critical temperature $T_c$ below its mean-field value $T_{\text{MF}}$. This occurs because of phase fluctuations. Coexistence of $d$-wave superconductivity with other phases in underdoped cuprates, such as antiferromagnetism (AF) or charge-density waves (CDW), may enhance the phase fluctuations and hence lower $T_c$.

% This may account for the lowering of $T_c$ just below optimal doping in electron-underdoped cuprates.

\end{abstract}
\maketitle

%%%%%%%%%%%%%%%%%%%%%%%%%%%%%%%%%%%%%%%%%%%%%%%% Introduction %%%%%%%%%%%%%%%%%%%%%%%%%%%%%%%%%%%%%%%%%%%%%%%%%%%%%%%%%%%%%%%%%%
%%

\section{Introduction}
\label{sec:Introduction}

Long-range order leads to emergent phenomena, such as generalized rigidities~\cite{Anderson}. The generalized rigidity associated with the superconducting state is superfluid stiffness, which allows phase coherence and supercurrent. Superfluid stiffness is accessible through penetration depth measurements~\cite{Uemura_similarities_1991}. In cuprates, one expects that strong interactions will modify the BCS predictions. Mott physics should make $\rho_s$ smaller than the BCS value as one approaches half-filling, but what about the effect of a competing order, such as antiferromagnetism (AF)~\citep{Mandal_e_doped_AFM_long_range_order,Motoyama2007_AFM_corr_more_prominent_e_doped,Armitage_Fournier_e_doped_RevModPhys_2010,saadaoui_phase_2015} or charge-density waves~\citep{Ghiringhelli_Tacon_2012,cyr-choiniere_CDW_sensitivity_2018}? This is particularly important for the electron-doped cuprates where long-range AF order has been observed far from half-filling, competing with $d$-wave superconductivity ($d$SC). This competition could explain the fall of both $H_{c_{2}}$ and $T_c$ in the underdoped cuprates~\cite{Mandal_e_doped_AFM_long_range_order,saadaoui_phase_2015,Keimer2015_nature}. 

%Despite the convincing experimental evidence linking $\rho_s$ and $T_c$ in underdoped cuprates, 
Very few theoretical works have addressed the question of the effect of microscopic homogeneous coexistence between AF and $d$SC on $\rho_s$. These works, based on mean-field calculations, have come to the conclusion that microscopic coexistence should decrease $\rho_s$~\citep{Micnas_Stiffness_2005,sharapov_superfluid_2006,atkinson_superfluid_2007,xiang_c_1996,mallik2018surprises}. Similar conclusions are reached with mean-field equations that use effective interactions generated by the functional renormalization group~\citep{Metzner_Stiffness:2019}. But all these theoretical works discard the effect of the strong electron-electron interaction and of the Mott transition, while it is known that the cuprates are doped Mott insulators~\cite{Anderson:1987}.

The best way to take Mott physics into account in two dimensions is to use cluster generalizations of dynamical mean-field theory~\citep{Maier_quantum_cluster_theories,KotliarRMP:2006,LTP:2006} for the Hubbard model. The only calculation of superfluid stiffness using these methods was done in the uniform superconducting state~\citep{Gull_Millis_interplane_cuprates_conductivity}, not in a phase where superconductivity coexists microscopically with antiferromagnetism. By microscopic coexistence, which we are interested in, we mean that both order parameters are present simultaneously and homogeneously in the ground state. By contrast, macroscopic coexistence would refer to what happens at a first-order transition where phases coexist in separate macroscopic regions. 

% The picture that emerges from the calculations of superconducting $T_c$ with cluster generalizations of dynamical mean-field theory that do not consider antiferromagnetism is that a) The Mott insulator at half-filling suffices to forbid superconductivity at half-filling. b) In four-site clusters, the superconducting dome is tilted towards half-filling contrary to what is observed in experiments~\citep{Fratino2016}. Increasing the cluster size to eight sites~\citep{Gull_Millis_interplane_cuprates_conductivity} and then twelve sites~\citep{Maier_Scalapino_2018}, the superconducting dome becomes more symmetric. This suggests that $T_c$ on small clusters detects only Cooper pair formation while the larger clusters are more sensitive to phase fluctuations that decrease $T_c$ in the underdoped regime. In fact, calculations of the pairing susceptibility in twelve sites clusters~\citep{Maier_Scalapino_2018} strongly suggests the importance of phase fluctuations in the underdoped regime. Since it is expected that the zero temperature $\rho_s$ is less dependent on the cluster size, the consistency of the whole picture can be checked by calculating this quantity in small clusters and verifying that its dependence on doping is similar to that observed in experiment. In addition, the value of $\rho_s$ at the lowest temperature gives an upper bound to $T_c$~\citep{Esterlis2018_Nature_Tc_bound}.

In this paper, we address the following questions: (1) What is the effect of the Mott transition on $\rho_s(n)$ near half-filling ($n=1$)? (2) Is microscopic coexistence with antiferromagnetism in the underdoped regime even more detrimental to $\rho_s$ than the Mott transition? (3) Is there a range of filling where BCS behavior is recovered? (4) What is the effect on $\rho_{zz}$ of the in-plane modulation of interplane hopping? To answer these questions, we compute the $c$-axis superfluid stiffness $\rho_{zz}$ for the one-band two-dimensional Hubbard model with band parameters appropriate to hole- and electron-doped cuprates. Along that direction, vertex corrections can be neglected, as we will discuss. We solve the Hubbard model using cellular dynamical mean-field theory (CDMFT) on a $2\times 2$ plaquette using an exact-diagonalization solver. The sites represent the Cu 3$d_{x^2-y^2}$ orbitals within the CuO$_2$ planes of the cuprates. We relax symmetries to allow both AF, $d$SC and their coexistence. We call ``coexistence'' the regime of calculations that allows for both the AF and $d$SC order parameters to coexist microscopically\footnote{Thereof, in the CMDFT calculations, the pure $d$SC phase can be emerging from pure or coexistence regimes.}. By contrast in the ``pure'' regime, the only symmetry breaking is $d$SC. The Hubbard model and the method to solve it is presented in \autoref{sec:Models_and_methods}. In \autoref{sec:superfluid_stiffness}, we follow up with the presentation of the formulae for $\rho_{zz}$ in both the pure $d$SC and coexisting AF+$d$SC states. We show the results in \autoref{sec:results}. The discussion is in \autoref{sec:discussion}. We conclude in \autoref{sec:conclusion}. The supplemental material~\footnote{See Supplemental Material at [URL will be inserted by publisher] for estimates of the in-plane superfluid stiffness and comments on the implications for $T_c$. Additional reference\cite{schrieffer2018theory}.} contains results for the in-plane superfluid stiffness, neglecting vertex corrections. That allows us to comment briefly in \autoref{sec:discussion} on the expected qualitative consequences of our results on the value of the superconducting transition temperature. This work is based on Ref.~\onlinecite{simard_master} where further details may be found. 

%%%%%%%%%%%%%%%%%%%%%%%%%%%%%%%%%%%%%%%%%%%%%%%% Model and Method %%%%%%%%%%%%%%%%%%%%%%%%%%%%%%%%%%%%%%%%%%%%%%%%%%%%%%%%%%%%%%
%%

\section{Model and method}
\label{sec:Models_and_methods}
The following subsections present in turn the model, the method and the periodization procedure. 
\subsection{Hubbard model}
\label{subsec:Hubbard_model}

To simulate interactions affecting electrons in high-T$_c$ cuprates, it was suggested by Anderson~\citep{Anderson:1987} that the Hubbard model 

\begin{align}
\label{eq:Hubbard_model_intro}
\hat{\mathcal{H}}=\sum_{ij,\sigma}t_{ij}\left(\hat{c}_{i,\sigma}^{\dagger}\hat{c}_{j,\sigma}+\hat{c}_{j,\sigma}^{\dagger}\hat{c}_{i,\sigma}\right)+U\sum_i\hat{n}_{i,\uparrow}\hat{n}_{i,\downarrow},
\end{align} 
would encompass  key aspects of these strongly correlated materials. Here, $t_{ij}$ are hopping amplitudes, $\sigma \in \{\uparrow,\downarrow\}$ are spin indices, $\hat{c}^{(\dagger)}_{i,\sigma}$ are annihilation (creation) operators in localized Wannier states labeled by $i,j$, while $\hat{n}_{i\sigma}=\hat{c}^{\dagger}_{i,\sigma}\hat{c}_{i,\sigma}$ is the number operator, and $U$ is the local repulsion normalized by the first-neighbor hopping term $t$.  The Hubbard model for CuO$_2$ planes of cuprates~\cite{Jorgensen_structural_properties_YBCO} is on a square lattice with spacing $a$. We take $c$ for the lattice spacing in the perpendicular $z$ direction. We set $\hbar$, $k_B$, electric charge $e$ and lattice spacings $a$, $c$ equal to unity for the figures. Physical units are restored for a few estimates and for some formulae. We used first-, second- and third-neighbor hopping terms to simulate bare electronic dispersion relations. We denote these hoppings respectively by $t$, $t'$ and $t''$ (see Figs.~\ref{fig:superfluid_stiffness:orbital_basis_spin_up} and \ref{fig:superfluid_stiffness:stacked_CuO2_planes_c_axis}). The tight-binding band parameters used are displayed in Table~\ref{table:ch_cuprates_supraconducteurs:motivation_theorique:table_tight_binding_parameters}~\citep{Band_structure_hopping_terms_Pavarini,kyung_pseudogap_2004}. YBa$_2$Cu$_3$O$_7$ (YBCO), La$_{2-x}$Sr$_x$CuO$_4$ (LSCO) and Bi$_2$Sr$_2$CaCu$_2$O$_{8+x}$ (BSCCO) are hole-doped compounds while Nd$_{2-x}$Ce$_x$CuO$_4$ (NCCO) is electron-doped. Nevertheless, to highlight the physics we consider the whole range of dopings for all sets of parameters.

In this work, we used the Green's functions obtained in Ref.~\onlinecite{foley_coexistence_2018} using CDMFT with the best available bath parametrization method, as described in the following subection. We also used the definitions given in Ref.~\onlinecite{foley_coexistence_2018} for the superconducting and antiferromagnetic order parameters, respectively denoted $\langle D\rangle$ and $\langle M\rangle$.

\begin{center}
 \begin{table}
    \begin{tabular}{|c|c|c|} \hline
        \hline
        compounds/parameters & $t^{\prime}/t$ & $t^{\prime \prime}/t$\\
        \hline
        YBCO/BSCCO & -0.3 & 0.2 \\
        LSCO/NCCO & -0.17 & 0.03 \\[1ex]
    
        \hline 
        \hline
    \end{tabular}       
    \caption{Tight-binding band parameters}
\label{table:ch_cuprates_supraconducteurs:motivation_theorique:table_tight_binding_parameters}
    \end{table}
    \end{center}

%\begin{table}[h!]
%\begin{center}

%\end{center}
%\end{table} 

\subsection{ED-CDMFT}
\label{subsec:CDMFT}

In CDMFT~\cite{Biroli_Kotliar_QCM_CDMFT}, a cluster of size $2\times 2$ representing a finite portion of the full lattice is hybridized to a bath of non-interacting electrons to simulate the effect of the environment on the cluster's electron Green's function. Hence, the number of orbitals with interactions is $N_c = 8$ (counting spin degeneracy). The cluster Hamiltonian $\hat{\mathcal{H}}^{\prime}$ including the hybridization to the baths reads~\cite{Senechal_quantum_cluster_methods,Caffarel:1994}

\begin{align}
\label{eq:Introduction:Impurity_Hamiltonian_CDMFT}
\hat{\mathcal{H}}^{\prime} =& -\sum_{ij,\sigma}t_{ij}\hat{c}_{i,\sigma}^{\dagger}\hat{c}_{j,\sigma}+U\sum_{i}\hat{n}_{i\uparrow}\hat{n}_{i\downarrow}\notag \\ 
&+\sum_{i\alpha,\sigma}\theta_{i\alpha,\sigma}\left(\hat{c}_{i,\sigma}^{\dagger}\hat{a}_{\alpha}+\text{H.c.}\right)+\sum_{\alpha,\sigma}\epsilon_{\alpha,\sigma}\hat{a}_{\alpha,\sigma}^{\dagger}\hat{a}_{\alpha,\sigma},
\end{align} where $\hat{c}^{(\dagger)}$ annihilates (creates) an electron on the cluster and $\hat{a}^{(\dagger)}$ annihilates (creates) an electron in the bath. The intra-cluster hopping matrix is $t_{ij}$ with $i$ and $j$ labelling the cluster sites and $\sigma\in \{\uparrow,\downarrow\}$. The baths are coupled to the cluster via the hybridization matrix $\theta_{i\alpha,\sigma}$ with $\alpha$ labelling the bath-orbital energy: the $\theta_{i\alpha,\sigma}$ represents the hopping of electrons between the cluster sites and the bath sites while $\epsilon_{\alpha,\sigma}$ is the energy of each orbital. The cluster Green's function is computed with an ED (impurity) solver based on the Lanczos algorithm \cite{Senechal_quantum_cluster_methods,Caffarel:1994}. 

In quantum cluster methods, the position is written as $\bm{r} = \bm{\tilde{r}}+\bm{R}$ where $\bm{\tilde{r}}$ is the base position of the cluster and $\bm{R}$ the position within the cluster. Likewise, a wave vector $\bm{k}$ in the Brillouin zone is decomposed as $\bm{k}=\bm{\tilde{k}}+\bm{K}$ where $\bm{\tilde{k}}$ belongs to the Brillouin zone of the superlattice of clusters (or reduced Brillouin zone) and $\bm{K}$ (which can also be seen as labeling the irreducible representations of the symmetry group of the cluster) belongs to the reciprocal superlattice.   

We work on the imaginary axis and the fermionic Matsubara frequencies are $\omega_n = 2\pi(n+1)/\beta$ where $n\in \mathbb{Z}$ and $\beta$ is the inverse temperature. The fictitious temperature defining the Matsubara grid is $\beta=50/t$.  The interacting cluster Green's function $\bm{\mathcal{G}}^{c,\sigma}_{\bm{R}_i\bm{R}_j}(i\omega_n)$ in the cluster-site mixed basis $(\bm{\tilde{k}},\bm{R})$ breaks down as follows

\begin{align}
\label{eq:Models_and_methods:ED_with_parametrisation:cluster_green_function_decomposition}
\bm{\mathcal{G}}^{c,\sigma}_{\bm{R}_i\bm{R}_j}(i\omega_n) = \left[(i\omega_n + \mu)\bm{I} - \bm{t^{\prime}} -\bm{\Gamma}^{\sigma}(i\omega_n)-\bm{\Sigma}_{c,\sigma}(i\omega_n)\right]_{ij}^{-1},
\end{align} 
where $\bm{\Sigma}_{c,\sigma}$ is the cluster self-energy matrix, $\bm{t^{\prime}}$ the intra-cluster hopping matrix, $\mu$ the chemical potential, and $\bm{\Gamma}^{\sigma}$ the hybridization function whose expression can be deduced from Eq.~\eqref{eq:Introduction:Impurity_Hamiltonian_CDMFT}:

\begin{align}
\label{eq:Models_and_methods:ED_with_parametrisation:hybridization_function}
\bm{\Gamma}^{\sigma}_{\bm{R}_i\bm{R}_j}\left(i\omega_n\right) = \sum_{\alpha}\frac{\theta_{i\alpha,\sigma}\theta_{j\alpha,\sigma}^{\ast}}{i\omega_n-\epsilon_{\alpha,\sigma}}.
\end{align} Each bath site is chosen to be in one of the irreducible representations of the cluster. That determines the symmetries of the $\theta$'s~\citep{foley_coexistence_2018}. In the following, on some occasions, the cluster-site indices and the spin will be left implicit. The $\bm{\Sigma}_{c,\sigma}$ used in our calculations is the one that satisfies the convergence criterion for the hybridization function. More specifically the cluster-projected Green's function in the cluster-site mixed basis $(\bm{\tilde{k}},\bm{R})$

\begin{align}
\label{eq:Models_and_methods:ED_with_parametrisation:projected_green_function_cluster}
\bm{\bar{\mathcal{G}}}^{\sigma}(i\omega_n) = \frac{N_c}{N}\sum_{\bm{\tilde{k}}}\frac{1}{i\omega_n + \mu -\bm{t}(\bm{\tilde{k}})-\bm{\Sigma}_{c,\sigma}(i\omega_n)}
\end{align} 
and the cluster Green's function Eq.~\eqref{eq:Models_and_methods:ED_with_parametrisation:cluster_green_function_decomposition} 
should be equal within a tolerance that sets the upper bound of the distance function $d$ which we minimize~\citep{foley_coexistence_2018}:

\begin{align}
\label{eq:Models_and_methods:ED_with_parametrization:distance_function}
d = \sum_{\substack{\bm{R}_i\bm{R}_j \\
                     i\omega_n\leq i\omega_c}} \sum_{\sigma}W(i\omega_n)\left|\left(\bm{\mathcal{G}}^{c,\sigma}(i\omega_n)^{-1}-\bm{\mathcal{\bar{G}}}^\sigma(i\omega_n)^{-1}\right)_{\substack{\bm{R}_i\bm{R}_j}}
                     \right|^2.
\end{align} When a finite number of bath orbitals is used to represent the environment, %the self-consistent relation \eqref{eq:Models_and_methods:ED_with_parametrization:distance_function} can't be exactly satisfied so 
one can't expect to obtain $d= 0$. Therefore, to capture the important degrees of freedom, one introduces a frequency cutoff $i\omega_c$, with $\omega_c = 2t$, to focus on the low-energy scale. The weight function $W(i\omega_n)$ is such that $W(i\omega_n)=1$ if $\omega_n < 2t$ and $W(i\omega_n)=0$ otherwise. Further details about the implementation can be found in Ref.~\onlinecite{foley_coexistence_2018}. In the equation for the lattice Green's function Eq.~\eqref{eq:Models_and_methods:ED_with_parametrisation:projected_green_function_cluster}, $\bm{t}(\bm{\tilde{k}}) = \bm{t^{\prime}}+\delta\bm{t}(\bm{\tilde{k}})$ represents the complete lattice hopping matrix, with $\delta\bm{t}(\bm{\tilde{k}})$ the intercluster hopping amplitude matrix carrying a phase proportional to both $\bm{\tilde{k}}$ and the lattice parameters. $N$ stands for the total number of sites on the full lattice. Once $d$ has been minimized, the full lattice Green's function $\bm{\mathcal{G}}(\bm{\tilde{k}},i\omega_n)$, dropping spin indices, reads

\begin{align}
\label{eq:Models_and_methods:ED_with_parametrization:lattice_Green_function}
\bm{\mathcal{G}}(\bm{\tilde{k}},i\omega_n)^{-1} = (i\omega_n + \mu)\bm{I} - \bm{t}(\bm{\tilde{k}})-\bm{\Sigma}_c(i\omega_n),
\end{align} where at each iteration the lattice self-energy is the same as that of the cluster $\bm{\Sigma}_c$:

\begin{align}
\label{eq:Models_and_methods:ED_with_parametrization:cluster_self_energy}
\bm{\Sigma}_c(i\omega_n) = (i\omega_n + \mu)\bm{I} - \bm{t^{\prime}} - \bm{\mathcal{G}}^c(i\omega_n)^{-1} - \bm{\Gamma}(i\omega_n).
\end{align} 

To account for superconductivity, the lattice Green's function Eq.~\eqref{eq:Models_and_methods:ED_with_parametrization:lattice_Green_function} is expressed in the following Nambu basis, assuming singlet pairing:

\begin{align}
\label{eq:Models_and_methods:ED_with_parametrization:cluster_site_basis_Green_function}
\begin{split}
&\hat{\Psi}_{\tilde{\bm{k}}}=\begin{pmatrix}
\hat{c}_{\bm{\tilde{k}}\uparrow,1}\\
\hat{c}_{\bm{\tilde{k}}\uparrow,2}\\
\vdots \\
\hat{c}^{\dagger}_{-\bm{\tilde{k}}\downarrow,N_c-1}\\
\hat{c}^{\dagger}_{-\bm{\tilde{k}}\downarrow,N_c}\\
\end{pmatrix}
\\
&\hat{\Psi}^{\dagger}_{\tilde{\bm{k}}}=\begin{pmatrix}
\hat{c}^{\dagger}_{\bm{\tilde{k}}\uparrow,1}&\hat{c}^{\dagger}_{\bm{\tilde{k}}\uparrow,2}&\hdots&\hat{c}_{-\bm{\tilde{k}}\downarrow,N_c-1}&\hat{c}_{-\bm{\tilde{k}}\downarrow,N_c}\\
\end{pmatrix}.
\end{split}
\end{align} 
In imaginary time, the definition is 
\begin{align}
\bm{\mathcal{G}}(\bm{\tilde{k}},\tau) = 
-\langle \hat{\mathcal{T}}_{\tau}\hat{\Psi}(\tau)\hat{\Psi}^{\dagger}(0)\rangle_{\hat{\mathcal{H}}}.
\end{align}
The above formulae for ED-CDMFT must be expressed in Nambu space, taking into account that they are no longer diagonal in Nambu indices. 

To avoid difficulties associated with the discreteness of the spectrum in ED, a ficticious temperature $\beta$ is introduced to compute $\rho_{zz}$. Since $\rho_{zz}$ converges rapidly with increasing $\beta$, this can be done with minimal effect on the accuracy of the zero-temperature calculation. All the results shown in \autoref{sec:results} were computed using $500$ Matsubara frequencies and $\beta = \frac{500}{t}$. In Ref.~\onlinecite{simard_master}, it is shown explicitly that $\rho_s$ converges fast with respect to the number of Matsubara frequencies used in the summation~($\rho_s\propto \frac{1}{(i\omega_n)^4}$) and with respect to the fictitious temperature.

\subsection{Periodization}
\label{subsubsec:Periodization}

\begin{figure}[h!] 
\center \includegraphics[clip=true,trim=0cm 0cm 0cm 0cm, width=0.8\columnwidth, height=0.8\columnwidth]{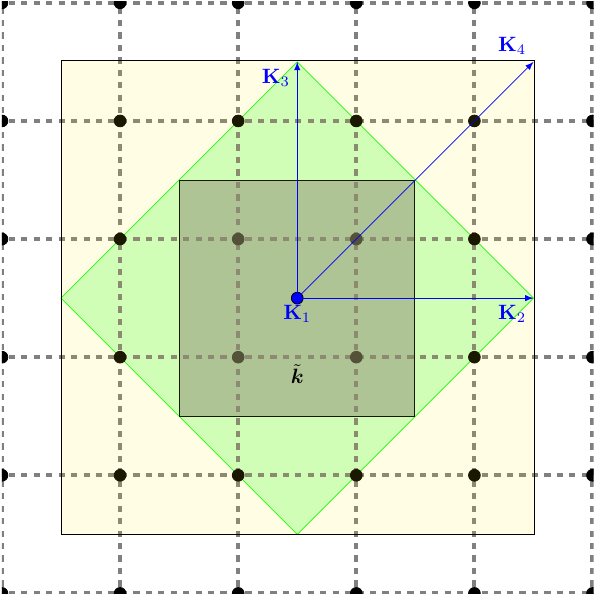}
\caption{The original Brillouin zone (BZ) is enclosed by the yellow square. The AF Brillouin zone (AF-BZ) is enclosed by the green diamond figure and the supercluster reduced Brillouin (rBZ) zone by the black square. $\bm{\mathcal{G}}(\bm{\tilde{k}},i\omega_n)$ is defined on the rBZ and has to be periodized to map onto the AF-BZ for the full Green's function $\mathcal{G}(\bm{k},i\omega_n)$ to have dimension $4\times 4$. In the case where there is only superconductivity, the wave vectors $\bm{K}_{i}$ with $i\in \{1,2,3,4\}$ are the reciprocal-superlattice wavevectors: $\bm{K}_1 = (0,0)$, $\bm{K}_2 = (\pi,0)$, $\bm{K}_3 = (0,\pi)$ and $\bm{K}_4 = (\pi,\pi)$.}
\label{fig:original_BZ_periodization}
\end{figure}

Once the lattice Green's function $\bm{\mathcal{G}}(\bm{\tilde{k}},i\omega_n)$ has been computed, one can periodize the latter to define it over the original Brillouin zone and recover translational invariance. For example, in a AF+$d$SC coexistence phase, periodizing $\bm{\mathcal{G}}(\bm{\tilde{k}},i\omega_n)$ to extend it over the reduced AF Brillouin zone (AF-BZ) seems natural (see Fig.~\ref{fig:original_BZ_periodization}). Doing so, the initially $8\times 8$ cluster Green's function in the mixed basis shrinks to $4\times 4$. The periodized cluster Green's function is~\cite{Periodization_procedure_Senechal,sakai_cluster_size_2012_Cum_per,Senechal:2015_Julich}

\begin{align}
\label{eq:cluster_green_function_periodization:cluster_green_function_periodization}
\bm{\mathcal{G}}(\bm{k},i\omega_n) = \frac{1}{N_c}\sum_{\bm{R}_i,\bm{R_j}}e^{-i\bm{k}\cdot\left(\bm{R}_i-\bm{R}_j\right)}\bm{\mathcal{G}}_{\bm{R}_i\bm{R}_j}\left(\bm{\tilde{k}},i\omega_n\right),
\end{align} 
where $N_c$ accounts for the number of cluster sites and $\bm{k}=\bm{\tilde{k}}+\bm{K}$. For periodization in the AF+$d$SC phase, $N_c=2$ and $\bm{K}_i\in \{(0,0),(\pi,0)\} \ \text{or} \ \{(0,0),(0,\pi)\}$, while in the SC state, $N_c=4$ and the reciprocal-superlattice wavevectors are $\bm{K}_i\in \{(0,0),(\pi,0),(0,\pi),(\pi,\pi)\}$. In the procedure with coexistence, the two sets of $\bm{K}$ values lead to exactly the same periodized Green's function, as can be understood with the aid of Fig.~\ref{fig:original_BZ_periodization}. Periodizing the Green's function Eq.~\eqref{eq:Models_and_methods:ED_with_parametrization:lattice_Green_function} using Eq.~\eqref{eq:cluster_green_function_periodization:cluster_green_function_periodization} reduces its dimensionality: for the case where AF and $d$SC coexist, the cluster Green's function in the reduced AF Brillouin zone shown in Fig.~\ref{fig:original_BZ_periodization} suffices to compute the superfluid stiffness. Eq.~\eqref{eq:cluster_green_function_periodization:cluster_green_function_periodization} is not a unitary transformation, because a unitary transformation would involve off-diagonal reciprocal-superlattice wavevectors, but one would not recover translational invariance.

From now on, we use the four-vector notation $k\equiv (\bm{k},i\omega_n)$ to lighten the notation, namely $\mathcal{G}(\bm{k},i\omega_n)\to \mathcal{G}(k)$. Note that cumulant periodization~\citep{Stanescu_Cumulant_2006} gives unphysical results for the superfluid stiffness~\cite{simard_master}, especially for YBCO-like tight-binding calculations. Self-energy periodization~\cite{Biroli_Kotliar_QCM_CDMFT} leads to unphysical states in the Mott gap at half-filling~\cite{Senechal:2015_Julich}, hence we do not consider these here. 

%The unitary transformation would be:
%
%\begin{align}
%\label{eq:cluster_green_function_periodization:cluster_green_function_periodization_unitary_transformation}
%&\bm{\mathcal{G}}(\bm{\tilde{k}}+\bm{K}_i,\bm{\tilde{k}}+\bm{K}_j)\notag\\ 
%&\phantom{\cdots}= \frac{1}{N_c}\sum_{\bm{R}_i\bm{R}_j}e^{-i\left(\bm{R}_i\cdot\bm{K}_i-\bm{R}_j\cdot\bm{K}_j\right)}e^{-i\bm{\tilde{k}}\cdot\left(\bm{R}_i-\bm{R}_j\right)}\bm{\mathcal{G}}_{\bm{R}_i\bm{R}_j}(\bm{\tilde{k}}).
%\end{align} 

%%%%%%%%%%%%%%%%%%%%%%%%%%%%%%%%%%%%%%%%%%%%%%%% Superfluid stiffness %%%%%%%%%%%%%%%%%%%%%%%%%%%%%%%%%%%%%%%%%%%%%%%%%%%%%%%%%%
%%

\section{Superfluid Stiffness}
\label{sec:superfluid_stiffness}

In this section we explain the general formalism for computing the superfluid stiffness on the superlattice of clusters and explain why vertex corrections can be neglected (\autoref{subsec:superfluid_stiffness:general_formulae}). Then we give the formula for pure $d$-wave superconductivity (\autoref{subsec:superfluid_stiffness:d_wave_superconductivity}), and finally the formula for the regime of coexistence between antiferromagnetism and superconductivity (\autoref{subsec:superfluid_stiffness:d_wave_superconductivity_AFM}). 

\subsection{General formula}
\label{subsec:superfluid_stiffness:general_formulae}

The superconducting order parameter is a consequence of spontaneous $U(1)$ symmetry breaking, the global phase being arbitrarily fixed. The phase rigidity, or superfluid stiffness, of the superconducting ground state accounts for the change in free energy when twisting the phase of the order parameter. In the linear response framework, the superfluid stiffness ${\rho}_{ab}$ for the response to a transverse vector potential $\bm{A}$ is related to the current-current correlation function by

\begin{align}
\label{eq:superfluid_stiffness:general_formulae:current_current_correlation_function}
{\rho}_{ab} &= \int_0^\beta d(\tau-\tau')\int d(\bm{r}-\bm{r}^\prime ) \langle \hat{\mathcal{T}}_{\tau}\hat{J}_a(\bm{r},\tau;\bm{A})\hat{J}_b(\bm{r}^{\prime},\tau';\bm{A})\rangle \notag \\
&= \frac{-1}{V}\int_0^\beta d\tau\int d(\bm{r}-\bm{r}^\prime ) \frac{\delta^2\mathcal{F}[\bm{\mathcal{G}}]}{\delta A_a(\bm{r},\tau)\delta A_b(\bm{r}^{\prime},0)}\bigg\rvert_{\bm{A}=\bm{0}},
\end{align} 
where subscripts $a,b\in \{x,y,z\}$ denote the cartesian axes, $V$ is the volume of a unit cell and $\mathcal{F}$ is the free energy (or energy at $T=0$) of the system. In particular, we evaluate the current $\hat{J}_z(\bm{r},\tau)$ along the $c$-axis   induced by a magnetic field applied in the transverse direction (in the plane). As we discuss below, this allows a calculation where neglecting vertex corrections can be justified. The vector potential $A_z(\bm{r}^{\prime},\tau^{\prime})$ representing the magnetic field is chosen along the $c$-axis as well. The above fomula Eq.~\eqref{eq:superfluid_stiffness:general_formulae:current_current_correlation_function} assumes that we are in the London limit where the kernel of the electromagnetic response can be evaluated in the zero wave vector ($\bm{q=0}$) limit so that linear response theory gives us
\begin{equation}
    \langle\hat{J}_a(\bm{q}=\bm{0},\omega=0)\rangle={\rho}_{ab}A_b(\bm{q}=\bm{0},\omega=0).
\end{equation}
Taking the curl of this equation and using Ampère's law $\nabla \times \bm{B}=\mu_0 \bm{J}$ with $\mu_0$ the  permeability of the vacuum, one finds that the London penetration depth $\lambda$ is related to the superfluid stiffness by 
\begin{align}
    \label{eq:London_penetration_length_superfluid_density_a}
    \lambda_{ab}^{-2} = {\rho}_{ab}\mu_0.
\end{align} 
In the BCS or Ginzburg-Landau formalism, this is written in terms of the superfluid density $n_s$ 
\begin{align}
    \label{eq:London_penetration_length_superfluid_density}
    \lambda_{ab}^{-2} = \frac{n_s e^2}{m^{\ast}}\mu_0,
\end{align}
 where $e$ is the electric charge and $m^{\ast}$ the effective mass of the electrons.

On the lattice, coupling of the Bloch electrons to the electromagnetic field is done via the Peierls substitution in the orbital basis ($\tilde{\bm{k}},\bm{R}$) (mixed basis). Since we can work in the $\bm{q=0}$ limit, the vector potential is a constant and the Peierls substitution leads to the replacement $\partial_{A_i} \rightarrow -\frac{e}{\hbar}\partial_{\bar{k}_i}$, where $\bar{k}_i\equiv k_i-\frac{e}{\hbar}A_i$, as long at the phase difference between atoms in the same unit cell is taken into account in the Fourier transforms~\citep{nourafkan_hall_2018}. Otherwise, the expression for the currents is different~\citep{Tomczak_Peierls:2009}. This is discussed further in \autoref{sec:appendice:Derivation_rho_AF+SC}. 

% Neglecting the vertex corrections is equivalent to neglecting the contribution of the Luttinger Ward functional  $\mathcal{F}[\bm{\mathcal{G}}] = -1/\beta\sum_{k}\ln{(-\bm{\mathcal{G}}(k)^{-1})}$ and expand $\bm{\rho}_{ab}$ in Eq.\eqref{eq:superfluid_stiffness:general_formulae:current_current_correlation_function} to obtain (restoring physical units) Commentaire: Dans ce cas la self-énergie n'apparaîtrait pas nulle part. Il faut justifier autrement.

When vertex corrections are neglected, the superfluid stiffness is given by
\begin{align}
\label{eq:superfluid_stiffness:general_formulae:superfluid_stiffness_expanded}
{\rho}_{ab} &= \frac{e^2}{\hbar^2\beta VN}\sum_{\bar{k},\sigma}\biggl(\text{tr}\left[\bm{\mathcal{G}}(\bar{k})\lambda^b_{\bar{\bm{k}}}T_{3(m\times m)}\bm{\mathcal{G}}(\bar{k})\lambda_{\bar{\bm{k}}}^aT_{3(m\times m)}\right]\notag\\ 
&+ \text{tr}\left[\bm{\mathcal{G}}(\bar{k})\lambda^{ab}_{\bar{\bm{k}}}\right]\biggr)\bigg\rvert_{\bm{A}=\bm{0}},
\end{align} 
where the trace $\text{tr}\left[\ldots\right]$ acts in the cluster-site mixed basis and $N$ is the number of unit cells. The first and second terms of Eq.~\eqref{eq:superfluid_stiffness:general_formulae:superfluid_stiffness_expanded} are, respectively, the paramagnetic and diamagnetic contributions to $\rho_{ab}$. Since the Nambu formalism involves a particle-hole transformation for the down electrons, we must evaluate the derivative with respect to the vector potential as follows $T_{0(m\times m)}\partial_{A_i} = -\frac{e}{\hbar}T_{3(m\times m)}\partial_{\bar{k}_i}$, where the tensors are defined by $T_{0(m\times m)}\equiv \sigma^0\otimes\mathbb{I}_{m\times m}$ and $T_{3(m\times m)}\equiv \sigma^3\otimes\mathbb{I}_{m\times m}$ with $\sigma^0$ the $2\times 2$ identity matrix and $\sigma^3$ the diagonal Pauli matrix whose components indicate whether we are in the spin up or spin-down part of the Nambu spinor Eq.~\eqref{eq:Models_and_methods:ED_with_parametrization:cluster_site_basis_Green_function}, a minus sign needing to be included in the spin-down part. The identity matrix $\mathbb{I}_{m\times m}$ depicts the remaining components of dimension $m\times m$ of the Nambu space. When vertex corrections are neglected, the partial derivative acts only on the kinetic energy term and not on the self-energy. Hence, we have defined the bare vertices
%%%%
\begin{align}\label{eq:vertices+}
    \lambda_{\bar{\bm{k}}}^i &\equiv \partial_{\bar{k}_i}\mathcal{H}_{\bar{\bm{k}},\sigma}^0\\
    \label{eq:vertices++}
    \lambda_{\bar{\bm{k}}}^{ji}&\equiv \partial_{\bar{k}_j}\partial_{\bar{k}_i}\mathcal{H}_{\bar{\bm{k}},\sigma}^0.
\end{align}

%Also, $\partial_{\bar{k}_i}\bm{\mathcal{G}}(\bar{k})^{-1}\equiv -\lambda_{\bar{\bm{k}}}^i$ and $\partial_{\bar{k}_j}\partial_{\bar{k}_i}\bm{\mathcal{G}}(\bar{k})^{-1}\equiv -\lambda_{\bar{\bm{k}}}^{ji}$. To obtain Eq.\eqref{eq:superfluid_stiffness:general_formulae:superfluid_stiffness_expanded}, we assumed that $\Sigma_c(\tilde{\bm{k}})$ does not couple to the vector potential $\bm{A}$, that is, $\partial_{\bar{k}_i}\mathcal{G}^{-1} = \partial_{\bar{k}_i}(i\omega_n + \mu - \mathcal{H}_{\bar{\bm{k}},\sigma}^0 - \Sigma_{\bm{k},\sigma}) = -\partial_{\bar{k}_i}\mathcal{H}_{\bar{\bm{k}},\sigma}^0 = -\lambda^i_{\bar{\bm{k}}}$. This is equivalent to dropping vertex corrections, as discussed in the next paragraph. 

The neglect of vertex corrections for $c-$axis superfluid stiffness is justified as follows. In the Green's function, the small $c$-axis hopping amplitude $t_{\perp}$ comes in the DMFT self-consistency equation only through the lattice Green's function, where it can be neglected compared with hopping in the plane. On the other hand, the current vertices coming from $t_{\perp}$ mean, in space-time, that the Green's functions entering the particle-hole bubble for the superfluid stiffness are initially in different planes; these two Green's functions do not allow hopping back to the same planes when $t_{\perp}$ is neglected. Since interactions are purely local in the Hubbard model, they can't act on Green's functions for electrons propagating in planes that are different. So the current vertex corrections for stiffness along the $c$-axis can be dropped out. The above argument also shows that the vertex corrections are of order $t_{\perp}^2$ compared with the leading terms. They can thus be neglected. Note that for the longitudinal response, which obeys the $f$-sum rule unlike the transverse response, the vector potential must be frequency dependent and, in addition, vertex corrections cannot be neglected~\citep{schrieffer2018theory}.  

The $c$-axis hopping amplitude branches out into many different forms depending on the class of cuprates studied. This is because the matrix element for hopping between planes depends a lot on which orbitals overlap. We defer for the details to Refs.~\onlinecite{chakravarty1993interlayer,xiang_c_1996,Panagopoulos_probing_c_axis_coupling_penetration_depth,Andersen1995,Markiewicz_hopping_tight_binding,Gull_Millis_interplane_cuprates_conductivity}. We chose a generic form describing $t_{\perp}$:

\begin{align}
\label{eq:superfluid_stiffness:general_formulae:bilayer_hopping_term}
t_{\perp}^2(\bm{k}) = t_{\text{bi}}^2\cos^2{k_z}\left(\cos{k_x}-\cos{k_y}\right)^4,
\end{align} 
where $t_{\text{bi}}\sim \frac{t}{25} = 10$meV~\citep{Markiewicz_hopping_tight_binding,Panagopoulos_probing_c_axis_coupling_penetration_depth}. For the figures, we take $t_{\text{bi}}=1$, except when we show values for the penetration depth in physical units. In momentum space, from ARPES experiments at temperatures between pseudogap crossover $T^{\ast}$ and $T_c$, the structure of the pseudogap appears to mimic the essential features of the $d$-wave superconducting gap~\cite{Loeser325_excitation_gap_noraml_state,Hoffman1148_quasiparticle_scattering}: the pseudogap is apparent only in the antinodal regions of the Brillouin zone where the $d$-wave gap is largest. Hence, the momentum dependence of $t_{\perp}$, of the form $(\cos{k_x}-\cos{k_y})^2$, suggests that the opening of the pseudogap in the CuO$_2$ plane will lead to a large effect on the superfluid stiffness. The current vertices $\lambda_{\bm{k}}^{i}$ in Eq.~\eqref{eq:superfluid_stiffness:general_formulae:superfluid_stiffness_expanded} are obtained from the partial derivative along $z$ of $t_{\perp}(\bm{k})$. 

%\begin{figure}[h!]
%\includegraphics[clip=true,trim=0cm 0cm 0cm 0cm, width=1.0\columnwidth, height=0.15\textwidth]{diagrammes_oli-crop.pdf}
%\caption{The paramagnetic and diamagnetic contributions to the superfluid stiffness $\bm{\rho}_{ab}$ (Eq.\eqref{eq:superfluid_stiffness:general_formulae:superfluid_stiffness_expanded}). The arrow lines represent the dressed electronic Green's functions in the 2d copper-oxide plane, $t_{\perp}$ is the current vertex along the c-axis and the label $i$ represents different layers in the unit cell. NOTE: Interplane hoppings are neglected in the propagation of the dressed in-plane electrons.}
%\label{fig:superfluid_stiffness_general_formulae:diagramatic_expansion}
%\end{figure}

 To compute the London penetration depth $\lambda_{c}\equiv \lambda_{zz}$ along the $c$-axis in physical units, we set nearest-neighbor in-plane hopping to $t\sim 250$meV, lattice constants to $a = b \simeq 3.8\angstrom$ and $c\simeq 11.7\angstrom$ for the YBCO-like results, $a = b \simeq 3.8\angstrom$ and $c\simeq 13.2\angstrom$ for the NCCO-like results with $t_{\text{bi}}\sim 10$meV. 
%~\citep{Jorgensen_structural_properties_YBCO}
\subsection{$d$SC regime}
\label{subsec:superfluid_stiffness:d_wave_superconductivity}

The superfluid stiffness without current vertex corrections comprising only $d$-wave superconductivity ($d$SC) reads~\citep{Gull_Millis_interplane_cuprates_conductivity,schrieffer2018theory}

\begin{align}
\label{eq:superfluid_stiffness:superfluid_stiffness_SC_only}
{\rho}_{zz}^{SC}=\frac{e^2}{\hbar^2\beta VN} &\sum_{k}\bar{t}_{\perp}^2(\bm{k})\times\notag\\
&\bigg(\text{tr}\left[\bm{\mathcal{G}}(k)\bm{\mathcal{G}}(k)\right]-\text{tr}\left[\sigma_3\bm{\mathcal{G}}(k)\sigma_3\bm{\mathcal{G}}(k)\right]\bigg),
\end{align} where $\sigma_3$ is the diagonal Pauli matrix. The trace $\text{tr}\left[\ldots\right]$ operates on Nambu space $\hat{\Psi}_{\bm{k}} = \left(\hat{c}_{\bm{k},\uparrow} \ \hat{c}^{\dagger}_{-\bm{k},\downarrow}\right)^{\intercal}$. The current vertices give a contribution

\begin{align}
\label{eq:superfluid_stiffness:d_wave_superconductivity_AFM:superluid_stiffness_current_vertex_c_axis}
\bar{t}^2_{\perp} &= \int_{-\pi}^{\pi}\frac{\mathrm{d}k_z}{2\pi}t^2_{\text{bi}}\sin^2{k_z}(\cos{k_x}-\cos{k_y})^4\notag\\
&= \frac{t^2_{\text{bi}}}{2}(\cos{k_x}-\cos{k_y})^4,
\end{align} 
where the integral over $k_z$ can be performed because $t{_\text{bi}}$ is neglected in the Green's functions. To compute $\bm{\rho}^{SC}_{zz}$ with the above formula, we first periodize the cluster Green's function $\bm{\mathcal{G}}(k)$ using the full set of superlattice reciprocal wavevectors $\bm{K}_i$ (see Fig.~\ref{fig:original_BZ_periodization}). The periodized Green's function is of size $2\times 2$. %It is important to note that the cluster Greens's function does not depend on $k_z$, but only on $\bm{k}_{\parallel}$.    

\begin{figure*}
    \includegraphics[clip=true,trim=0cm 0cm 0cm 0cm, width=\columnwidth ]{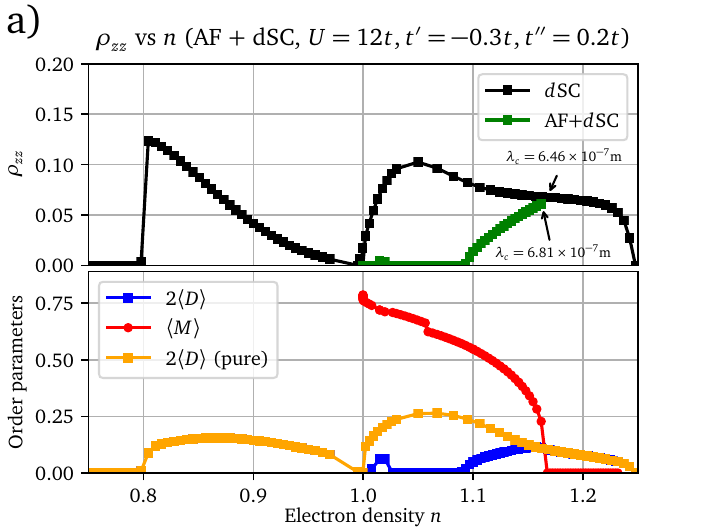}
    \includegraphics[clip=true,trim=0cm 0cm 0cm 0cm,width=\columnwidth ]{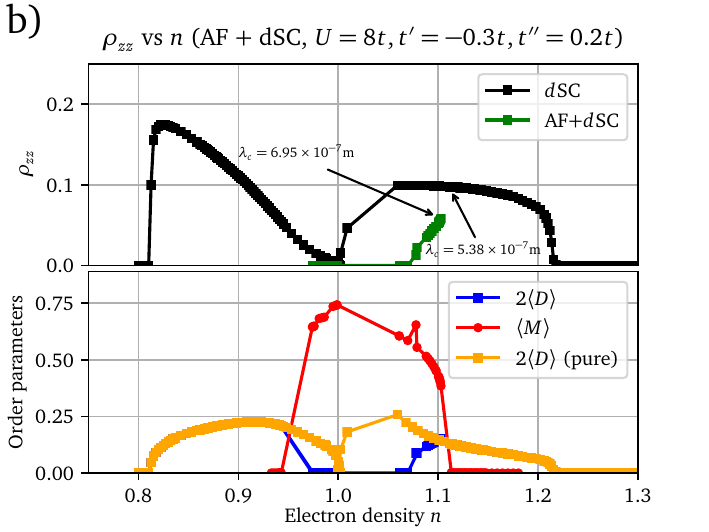}
    \caption{a) $\rho_{zz}$ as a function of band filling $n$ in both pure $d$SC and coexisting AF+$d$SC states for $U=12t$, $t^{\prime}=-0.3t$ and $t^{\prime\prime} = 0.2t$. The green squares show $\rho^{AF+SC}_{zz}$ in coexistence regime, and the black squares show $\rho^{SC}_{zz}$ computed in the pure regime (with coexistence forbidden). The calculated values of $\lambda_c$ in physical units are of the same order of magnitude as experimental measurements of the $c$-axis superfluid penetration depth in hole-doped compounds~(Ref.~\onlinecite{Panagopoulos_probing_c_axis_coupling_penetration_depth}). The bottom subfigure illustrates the $d$SC order parameter in the pure regime $\langle D\rangle\ (\text{pure})$ (orange), and the $d$SC $\langle D\rangle$ (blue) and AF $\langle M\rangle$ (red) order parameters as a function of $n$ in the coexistence regime. b) $\rho_{zz}$ as a function of the electron density $n$ in both pure $d$SC and microscopic AF+$d$SC states for $U=8t$, $t^{\prime}=-0.3t$ and $t^{\prime\prime}=0.2t$. The symbols are the same as in Fig.~\ref{fig:superfluid_stiffness_coexistence_U12_U8_tperp}~a).}
    \label{fig:superfluid_stiffness_coexistence_U12_U8_tperp}
\end{figure*}

\subsection{Coexistence regime $d$SC + AF}
\label{subsec:superfluid_stiffness:d_wave_superconductivity_AFM}

We derived a formula to compute the superfluid stiffness in the regime where $d$-wave superconductivity and antiferromagnetism coexist homogeneously. This formula requires that one periodizes the cluster Green's function to map onto the reduced AF Brillouin zone (AF-BZ). 

First we define 
\begin{align}
\label{eq:superfluid_stiffness:d_wave_superconductivity_AFM:T_matrices_representation}
T_{lm}=\sigma^l_{\alpha\beta}\tau^m_{ab},
\end{align} 
where $\sigma$ and $\tau$ are Pauli matrices,  $\sigma^l$ acting in Nambu space and the $\tau^m$ acting in the AF sublattice space spanned by sublattices $A$ and $B$. We define $\sigma_0$ and $\tau_0$ as the identity matrix $\mathbb{I}_{2\times 2}$. The superfluid stiffness when AF and $d$SC coexist then reads ($\bm{\mathcal{G}}(k)\to \bm{\mathcal{G}}$):

\begin{align}
\label{eq:superfluid_stiffness:d_wave_superconductivity_AFM:superluid_stiffness_periodization}
{\rho}_{zz}^{AF+SC}&=\frac{e^2}{\hbar^2\beta VN}\sum_{k}\bar{t}_{\perp}^2(\bm{k})\times\notag\\
&\bigg(\text{tr}\left[\bm{\mathcal{G}}T_{01}\bm{\mathcal{G}}T_{01}\right]-\text{tr}\left[\bm{\mathcal{G}}T_{31}\bm{\mathcal{G}}T_{31}\right]\bigg).
\end{align} A detailed derivation of Eq.~\eqref{eq:superfluid_stiffness:d_wave_superconductivity_AFM:superluid_stiffness_periodization} is given in Appendix~\ref{sec:appendice:Derivation_rho_AF+SC}. It can be extended to any phase coexistence scenario.  In the above equation, the Green's functions extracted from the CDMFT procedure are periodized using Eq.~\eqref{eq:cluster_green_function_periodization:cluster_green_function_periodization} with either $\bm{K}_y = \{(0,0),(0,\pi)\}$ or $\bm{K}_x = \{(0,0),(\pi,0)\}$ as the set of superlattice wavevectors~(\textit{cf.}~Fig.\ref{fig:original_BZ_periodization}).  The Green's functions are then of dimension $4\times 4$ instead of $8\times 8$. Note that the dimension of the Green's function in Eq.~\eqref{eq:Models_and_methods:ED_with_parametrization:cluster_site_basis_Green_function} is $8\times 8$ for unperiodized Green's functions.

It is important to stress that the Green's functions must be periodized in the AF-BZ prior to using Eq.~\eqref{eq:superfluid_stiffness:d_wave_superconductivity_AFM:superluid_stiffness_periodization} whether they have converged to a pure $d$SC state, a pure AF state, or a microscopic AF+$d$SC state. When the solution converges to a pure $d$SC state instead of microscopic AF+$d$SC, the superfluid stiffness obtained with either periodizations, namely Eq.~\eqref{eq:superfluid_stiffness:superfluid_stiffness_SC_only} or~\eqref{eq:superfluid_stiffness:d_wave_superconductivity_AFM:superluid_stiffness_periodization}, are indistiguishable on the plots.

%%%%%%%%%%%%%%%%%%%%%%%%%%%%%%%%%%%%%%%%%%%%%%%% Results %%%%%%%%%%%%%%%%%%%%%%%%%%%%%%%%%%%%%%%%%%%%%%%%%%%%%%%%%%%%%%%%%%%
%%

\section{Results}
\label{sec:results}

We study the superfluid stiffness $\rho_{zz}$ for a variety of parameters within the one-band Hubbard model Eq.~\eqref{eq:Hubbard_model_intro}, both with and without homogeneous microscopic AF+$d$SC coexistence. We find such coexistence in the CDMFT solutions of the cluster Green's function only on the electron-doped side $(n > 1)$ . The hole-doped side corresponds to band filling $n<1$. Whether antiferromagnetism is present or not, superconductivity is supressed at half-filling when the Hubbard interaction $U$ becomes larger than the value $U_c~\sim 6$ that leads to a Mott insulator (see Figs.~\ref{fig:superfluid_stiffness_coexistence_U12_U8_tperp}~a), \ref{fig:superfluid_stiffness_coexistence_U12_U8_tperp}~b) and \ref{fig:superfluid_stiffness_coexistence_U655_U5_tperp}~a)). Overdoping means small $n$ for $n<1$ and large $n$ for $n>1$. In both cases, underdoping is near $n=1$. 

We consider in turn band parameters that are close to those of YBCO and those of NCCO. The last subsection will show the effect of the $\bm{k}_{\parallel}$-dependence of $t_{\perp}$ in Eq.~\ref{eq:superfluid_stiffness:general_formulae:bilayer_hopping_term}, giving us some insight on the parts of the Fermi surface that are most relevant for superconductivity. 

\subsection{YBCO-like band parameters}

Figures \ref{fig:superfluid_stiffness_coexistence_U12_U8_tperp}~a) and \ref{fig:superfluid_stiffness_coexistence_U12_U8_tperp}~b) illustrate both $\rho^{SC}_{zz}$ and $\rho^{AF+SC}_{zz}$ with respect to band filling $n$ per Cu 3$d_{x^2-y^2}$ orbital for the YBCO tight-binding parameters at $U=12t$ and $U=8t$, respectively. 

The superfluid stiffness for both values of $U$ and for both hole- and electron-doping falls abruptly to zero in the overdoped regimes, where there is no coexistence. This suggests that in this limit, the system eventually reaches BCS-like behavior where at $T=0$ that sudden drop is expected. Finite resolution in the distance function, which contains an artifical temperature, probably explains why that drop is not perfectly discontinuous.

\begin{figure*}
    \includegraphics[clip=true,trim=0cm 0cm 0cm 0cm, width=\columnwidth]{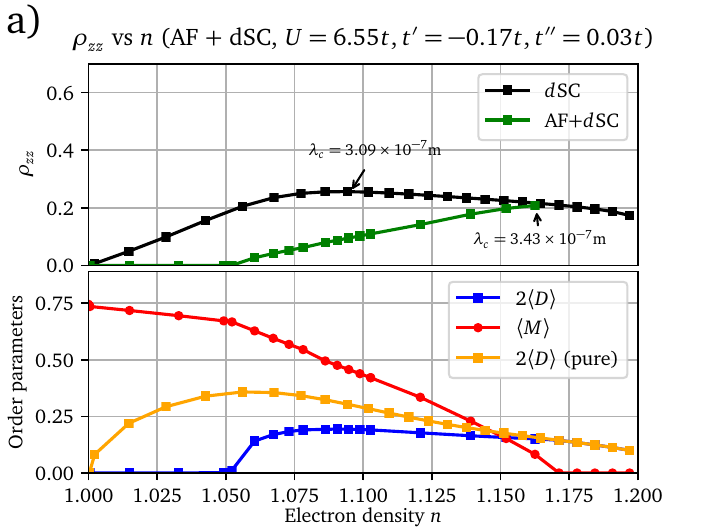}
    \includegraphics[clip=true,trim=0cm 0cm 0cm 0cm, width=\columnwidth]{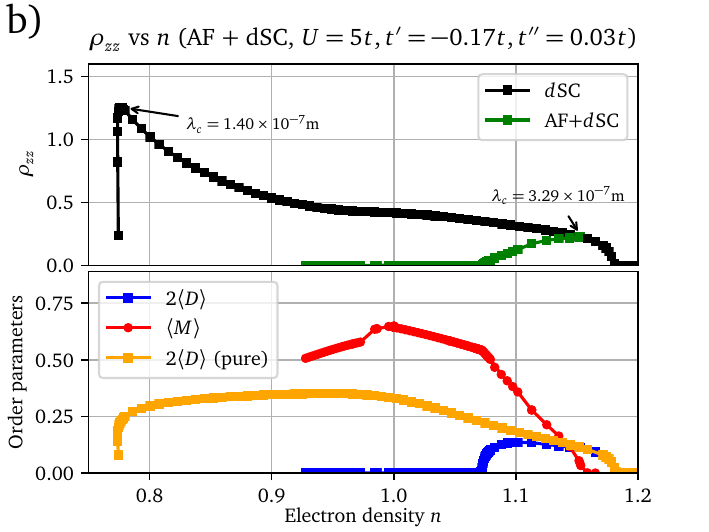}
    \caption{a) $\rho_{zz}$ as a function of band filling $n$ in both pure $d$SC and AF+$d$SC states for $U=6.55t$, $t^{\prime}=-0.17t$ and $t^{\prime\prime}=0.03t$. The symbols are defined in Fig.~\ref{fig:superfluid_stiffness_coexistence_U12_U8_tperp}. b) $\rho_{zz}$ as a function of band filling $n$ in both pure $d$SC and AF+$d$SC states for $U=5t$, $t^{\prime}=-0.17t$ and $t^{\prime\prime}=0.03t$. The symbols are defined in Fig.~\ref{fig:superfluid_stiffness_coexistence_U12_U8_tperp}.}
    \label{fig:superfluid_stiffness_coexistence_U655_U5_tperp}
\end{figure*}

Contrast this BCS-like behavior with the behavior near half-filling for $U=12$ in Fig.~\ref{fig:superfluid_stiffness_coexistence_U12_U8_tperp}~a) where the fall is much more gradual, as has been observed experimentally both along the $c$-axis and in the plane~\cite{Panagopoulos_probing_c_axis_coupling_penetration_depth,Uemura_T_c_scaling_rho_s_several_layers_1989,Tallon_T_c_scaling_rho_s_2003}. This is clearly the effect of the Mott transition since it does not appear when $U$ is not large enough ($U<U_c$), as can be seen in Fig.~\ref{fig:superfluid_stiffness_coexistence_U655_U5_tperp}~b). The gradual fall of the superfluid stiffness has been interpreted as indicating that the superfluid density, as defined by the penetration depth Eq.~\eqref{eq:London_penetration_length_superfluid_density}, vanishes at half-filling and increases roughly proportionally to the doping, as if the number of carriers had to be measured with respect to half-filling.

Let us move to the effect of the competition with antiferromagnetism. Although there is no coexistence on the hole-doped side, antiferromagnetism is detrimental to superconductivity for $U=8$, as can be seen in Fig.~\ref{fig:superfluid_stiffness_coexistence_U12_U8_tperp}~b) where the superconducting order parameter vanishes before half-filling is reached.  

Comparing figures~\ref{fig:superfluid_stiffness_coexistence_U12_U8_tperp}~a) and \ref{fig:superfluid_stiffness_coexistence_U12_U8_tperp}~b) in the region where there is coexistence, namely on the electron-doped side, we see that as $U$ increases, the domain of dopings where $d$SC and AF coexist grows. With increasing $U$,  coexistence also ends at larger dopings when it reaches the pure superconducting phase near optimal doping. The jump in superfluid stiffness at this point is quite remarkable, but it may just reflect that the transition between the pure and coexisting phases is a first-order transition. The numerical values of $c$-axis superfluid stiffness given on the plots in physical units are of the correct order of magnitude compared with experimental measurements in cuprates~\cite{Panagopoulos_probing_c_axis_coupling_penetration_depth}. Another remarkable property of superfluid stiffness in the coexistence region is that it is small and it decreases extremely rapidly as half-filling is approached. Coexistence ends relatively far away from half-filling.

In the coexistence regime, the converged cluster Green's functions were periodized in the AF-BZ and used in Eq.~\eqref{eq:superfluid_stiffness:d_wave_superconductivity_AFM:superluid_stiffness_periodization}. On the other hand, the superfluid stiffness in the pure $d$SC phase was computed by periodizing the cluster Green's function in the BZ and used in Eq.~\eqref{eq:superfluid_stiffness:superfluid_stiffness_SC_only}. The agreement between both formulae when there is only a pure $d$SC phase (not shown on the figures) is non trivial and suggests that the superfluid stiffness formulae and the methods are consistent. This correspondence is also observed for calculations with the NCCO-like parameters.

Electron-doped material generally do not have band parameters close to those of YBCO. Electron-doped NCCO-like band parameters are explored in the next subsection.

%\onecolumngrid

% \begin{widetext}
    
%     \begin{figure}[h]
%         \centering
%         \begin{minipage}[t]{.48\textwidth}
%           %\centering
%           \includegraphics[clip=true,trim=0cm 0cm 0cm 0cm, width=\linewidth, height=\textwidth]{Figure_U655_tperp_Python.pdf}
%           \captionof{figure}{$\rho_{zz}$ as a function of band filling $n$ in both $d$SC-only and AF+$d$SC coexistence states for $U=6.55t$, $t^{\prime}=-0.17t$ and $t^{\prime\prime}=0.03t$. The symbols are defined in Fig.~\ref{fig:superfluid_stiffness_coexistence_U12_tperp}}%{}
%           \label{fig:superfluid_stiffness_coexistence_U655_tperp}
%         \end{minipage}%
%         \begin{minipage}[t]{.48\textwidth}
%           %\centering
%           \includegraphics[clip=true,trim=0cm 0cm 0cm 0cm, width=\linewidth, height=\textwidth]{Figure_U5_tperp_Python.pdf}
%           \captionof{figure}{$\rho_{zz}$ as a function of band filling $n$ in both $d$SC-only and AF+$d$SC states for $U=5t$, $t^{\prime}=-0.17t$ and $t^{\prime\prime}=0.03t$. The symbols are defined in Fig.~\ref{fig:superfluid_stiffness_coexistence_U12_tperp}: black stars are for $d$SC-only phase computed in the pure regime while green stars are for the microscopic AF+$d$SC phase.}%{}
%           \label{fig:superfluid_stiffness_coexistence_U5_tperp}
%         \end{minipage}
%     \end{figure}

% \end{widetext}

%\twocolumngrid

\subsection{NCCO-like band parameters}

Comparing calculations with experiments suggests that electron-doped cuprates, such as NCCO, are described by a Hubbard model with a value of $U$ in the vicinity of the Mott transition~\cite{senechal_hot_2004,kyung_pseudogap_2004,LTP:2006,weber_strength_2010}. The results for $\rho_{zz}$ appear in Figs.~\ref{fig:superfluid_stiffness_coexistence_U655_U5_tperp}~a) and \ref{fig:superfluid_stiffness_coexistence_U655_U5_tperp}~b). Contrary to above, the discontinuity in $\rho_{zz}$ when antiferromagnetism appears near optimal doping has disappeared. The values of $U$ are quite close for the two plots, $U=6.55t$ in Fig.~\ref{fig:superfluid_stiffness_coexistence_U655_U5_tperp}~a) and $U=5t$ in Fig.~\ref{fig:superfluid_stiffness_coexistence_U655_U5_tperp}~b), leading to values of $\rho_{zz}$ that are quite close on the electron-doped side near optimal doping. While $\rho_{zz}$ looks continuous as a function of $n$ in Fig.~\ref{fig:superfluid_stiffness_coexistence_U655_U5_tperp}~b) when antiferromagnetism appears upon decreasing doping, in the doped Mott insulator regime (see Fig.~\ref{fig:superfluid_stiffness_coexistence_U655_U5_tperp}~a)), there is a rapid change in slope as a function of $n$ when antiferromagnetism appears.  
%Note that the YBCO tight-binding band parameters provide better nesting of the two AF sub-bands. This better nesting must favor the apparent first-order like behavior observed for $\langle M\rangle$ near optimal doping on the electron-doped side. 

%We can't really comment with certainty on the nature of the phase transition associated with the apparition of the antiferromagnetism ($\langle M\rangle$). We would need to zoom in on the doping range in the vicinity where $\langle M\rangle$ becomes finite and increase the cluster size, in order to rule on the nature of the phase transition. Although it is not clear from the AF order parameter whether AF appears in a first order fashion or not, the jump in $\rho_{zz}$ suggests that it does.

\begin{figure*}
    \includegraphics[clip=true,trim=0cm 0cm 0cm 0cm, width=\columnwidth]{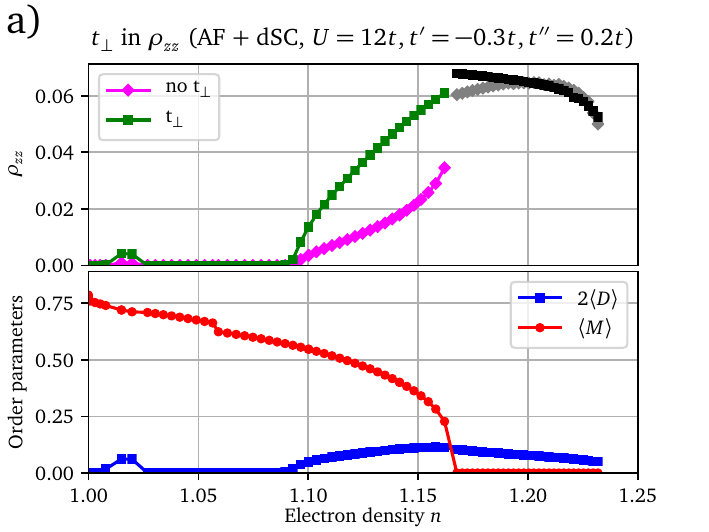}
    \includegraphics[clip=true,trim=0cm 0cm 0cm 0cm, width=\columnwidth]{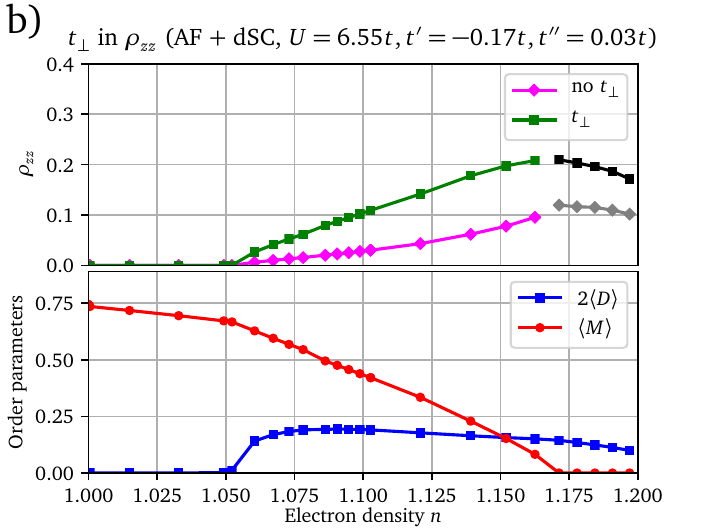}
    \caption{a) Effect of the $\bm{k}_{\parallel}$-dependence of $t_{\perp}$ on $\rho^{AF+SC}_{zz}$ in the coexistence regime for $U=12t$, $t^{\prime}=-0.3t$ and $t^{\prime\prime}=0.2t$. Only the electron-doped side is shown. The green and black squares are for $\rho_{zz}$ as a function of $n$ with the $\bm{k}_{\parallel}$-dependence of $t_{\perp}$ included: these results are taken from Fig.~\ref{fig:superfluid_stiffness_coexistence_U12_U8_tperp}~a). The magenta (grey) diamonds on the other hand show $\rho^{AF+SC}_{zz}$ with (without) coexistence as a function of $n$ replacing the $\bm{k}_{\parallel}$-modulation of $t_{\perp}$ in Eq.~\eqref{eq:superfluid_stiffness:d_wave_superconductivity_AFM:superluid_stiffness_periodization} by its reduced-Brillouin-zone average. Thus, the grey diamonds show the effect of a missing $t_{\perp}$ when $\langle M\rangle$ vanishes and $\langle D\rangle$ dominates in the case where both order parameters are allowed in the calculations. The bottom subfigure illustrates the AF and $d$SC order parameter amplitudes, $\langle M\rangle$ and $\langle D\rangle$ respectively, as a function of $n$.  b) $\rho_{zz}$ for $U=6.55t$, $t^{\prime}=-0.17t$ and $t^{\prime\prime}=0.03t$. The symbols have the same meaning as in Fig.~\ref{fig:superfluid_stiffness_coexistence_U12_U655_tperp_vs_no_tperp}~a). The green and black squares are taken from Fig.~\ref{fig:superfluid_stiffness_coexistence_U655_U5_tperp}~a).}
    \label{fig:superfluid_stiffness_coexistence_U12_U655_tperp_vs_no_tperp}
\end{figure*}

Even though the values of $U$ in Figs.~\ref{fig:superfluid_stiffness_coexistence_U655_U5_tperp}~a) and \ref{fig:superfluid_stiffness_coexistence_U655_U5_tperp}~b) are quite close, the difference between the two is quite striking. The case $U=5t$ in Fig.~\ref{fig:superfluid_stiffness_coexistence_U655_U5_tperp}~b) is below $U_c$ for the Mott transition. This allows superconductivity to survive at half-filling when we do not allow antiferromagnetism to set in. The fall of $\rho_{zz}$ in the two overdoped regimes is abrupt, in BCS-like fashion. BCS would predict that $\rho_{zz}$ is proportional to band filling. Since the non-interacting Fermi surface is hole-like even on the electron-doped side, this is consistent with the increase in superfluid stiffness as $n$ decreases or hole density $\left| 1-n\right|$ increases. The non-interacting van-Hove singularity where the Fermi surface becomes electron-like is at $n=0.8$, but this is shifted by interactions. 

\subsection{Effect of the ${k}_{\parallel}$-dependence of $t_{\perp}$}

Since $t_{\perp}$ is maximum at the $(\pi,0)$, $(0,\pi)$ points, as can be seen from $t_{\perp}^2(\bm{k}) = t_{\text{bi}}^2\cos^2{k_z}\left(\cos{k_x}-\cos{k_y}\right)^4$ (Eq.~\eqref{eq:superfluid_stiffness:general_formulae:bilayer_hopping_term}), an interesting question arises. Since both the pseudogap and the antiferromagnetic gap vary along the Fermi surface in the plane, their effect on $c$-axis superfluid stiffness $\rho_{zz}$ should be influenced by modulations of the $c$-axis hopping integral in the plane. What is the net effect of this modulation? The answer is in Figs.~\ref{fig:superfluid_stiffness_coexistence_U12_U655_tperp_vs_no_tperp}~a), \ref{fig:superfluid_stiffness_coexistence_U12_U655_tperp_vs_no_tperp}~b).

We have computed $\rho_{zz}$ with and without the $\bm{k}_{\parallel}$-dependence of $t_{\perp}$. In the plots, by ``no $t_{\perp}$'', we mean ``in the absence of the $\bm{k}_{\parallel}$-dependence of the bilayer hopping term $t_{\perp}$''. In other words, we have replaced the in-plane modulation of perpendicular hopping $(\cos{k_x}-\cos{k_y})^4$ by $9/8$ since this is its average over the AF Brillouin zone. Figure \ref{fig:superfluid_stiffness_coexistence_U12_U655_tperp_vs_no_tperp}~a) shows the effect of the $\bm{k}_{\parallel}$-dependence on $\rho_{zz}$ for YBCO band parameters, $U=12t$ and $n>1$. Figure \ref{fig:superfluid_stiffness_coexistence_U12_U655_tperp_vs_no_tperp}~b) shows the same for NCCO band parameters, $U=6.55t$ and $n>1$. The results are qualitatively similar for the two sets of parameters. 

The $\bm{k}_{\parallel}$-dependent term of $t_{\perp}$ (Eq.~\eqref{eq:superfluid_stiffness:general_formulae:bilayer_hopping_term}) takes its maximum values in the portions of the Brillouin zone where $\bm{k}=(0,\pi)$ or $(\pi,0)$. These portions of the BZ are the antinodal regions where the $d$SC gap is the largest. Hence this is the region of the Brillouin zone that contributes most to the superfluid stiffness. In the coexistence region, which is electron-doped, the AF Fermi surface still has weight where the superconducting gap is largest. Hence, increasing the importance of these regions makes the superfluid stiffness larger. Also, the $\mathbf{k}_{\parallel}$-dependence of $t_{\perp}$ increases the contribution to $\rho_{zz}$ of the states in the vicinity of the van-Hove singularity on the hole-doped side (not shown). The latter can be checked by means of simple mean-field calculations. 

By contrast, when superconductivity gaps the pseudogap normal state without coexisting antiferromagnetism, the situation is different. The pseudogap in the normal state is near $(\pm\pi/2,\pm\pi/2)$ in the electron-doped case. The superconductivity in that region is effective in lowering the energy because it replaces the pseudogap by quasiparticles. Hence, a more uniform weighting of the contributions across the Brillouin zone is more favourable in this case (not shown). This is also why the superfluid stiffness becomes larger without the $\mathbf{k}_{\parallel}$-dependence for $n>1.2$ in Fig.~\ref{fig:superfluid_stiffness_coexistence_U12_U655_tperp_vs_no_tperp}~a).  

% In the regime of coexistence, the removal of the $\bm{k}_{\parallel}$-dependence is even more remarkable: it is because when developing $\left(\cos{k_x}-\cos{k_y}\right)^2 = 1 - 2\cos{k_x}\cos{k_y} + \left[\cos{2k_x}+\cos{2k_y}\right]/2$, one stumbles upon the first-, second- and third-nearest neighbor hopping terms from one plane to another. Therefore, removing that dependence, one drops the second- and third-nearest neighbor contributions to $\rho_{zz}$~(see Fig.\ref{fig:superfluid_stiffness:stacked_CuO2_planes_c_axis}).

%%%%%%%%%%%%%%%%%%%%%%% Discussion %%%%%%%%%%%%%%
%%%%%%%%%%%%%%%%%%%

\section{Discussion}
\label{sec:discussion}

The $c$-axis striffness $\rho_{zz}$ has been calculated in Ref.~\onlinecite{Gull_Millis_interplane_cuprates_conductivity} using 8-site DCA for $U=6t>U_c$ and $\beta=60/t$ for particle-hole symmetric in-plane nearest-neighbor hopping. Their conclusions are qualitatively similar to the ones shown in Fig.~\ref{fig:superfluid_stiffness_coexistence_U12_U8_tperp}~b) and \ref{fig:superfluid_stiffness_coexistence_U655_U5_tperp}~a): Mott physics suffices to lead to a vanishing superfluid stiffness as half-filling is approached and the fall of $\rho_{zz}$ when superconductivity disappears is more BCS-like in the overdoped regime. The authors noted that finite-temperature effects were likely to influence the results in the latter case, as also suggested in Ref.~\onlinecite{Metzner_Stiffness:2019}.

As noted in the introduction, due to the finiteness of the cluster, the order parameter $\langle D\rangle$ indicates the formation of local Cooper pairs and the order parameter $\langle M\rangle$ indicates the formation of $(\pi,\pi)$ particle-hole bound states, or equivalently, local AF spin correlations. 

%%%%%%%%%%%%%%%%%%%%%%%%%%%%%%%%%%%%%%%%%%%%%%%%  %%%%%%%%%%%%%%%%%%%%%%%%%%%%%%%%%%%%%%%%%%%%%%%%%%%%%%%%%%%%%%%%%%
%%

\subsection{Relation between $\rho_s$ and $T_c$}
\label{sec:Tc}

Before we discuss our results further, we comment on the possible relation between $\rho_s$ and $T_c$.
Since cuprate superconductors are layered highly anisotropic superconductors, one expects Kosterlitz-Thouless physics~\citep{nelson_universal_1977,Kosterlitz_Thouless_2D_topological_phase_transition}
to play an important role in determining the actual superconducting transition temperature $T_c$~\citep{Emery1995_importance_of_phase_fluctuations,lee_theory_2008}. This is brought out by experiments~\citep{boschini_collapse_2017,Yong_McCray_Lemberger_Naamneh_Kanigel_Randeria_2012,Hetel_Lemberger_Randeria_2007} even though there are contradictory views~\citep{Range_SC_fluct:2011,chang_gaussian_fluctuations:2012,Tafti_Nernst_Phase_2014}. The relevance of Kosterlitz-Thouless physics comes out clearly in the experimentally-determined proportionality between $T_c$ and $\rho_s$ in underdoped cuprates, both hole-doping\citep{Uemura_T_c_scaling_rho_s_several_layers_1989,Tallon_T_c_scaling_rho_s_2003} and electron-doping~\citep{li_hole-pocket-driven_2018}. 

Finite-temperature DCA studies with eight sites~\citep{Gull_Millis_interplane_cuprates_conductivity} and twelve sites~\citep{Maier_Scalapino_2018} show that in the finite-temperature underdoped regime, increased phase fluctuations improve the agreement between the calculated and the observed shape of the superconducting transition-temperature dome. These theoretical results are consistent with the importance of long wave-length phase fluctuations, as in the Kosterlitz-Thouless theory. 

When Kosterlitz-Thouless physics applies, $T_c$ can be bounded~\citep{Emery1995_importance_of_phase_fluctuations,Esterlis2018_Nature_Tc_bound} using the zero-temperature value of $\rho_s$. However, it is the in-plane $\rho_s$ that counts. In the supplemental material~\citep{Note2}, we compute that quantity, assuming that vertex corrections can be neglected.  We find that the zero-temperature in plane $\rho_s$ scales with doping in a manner very similar to the $c$-axis stiffness. Further finite-temperature calculations and their relation to $T_c$ will be the subject of future work.    

\subsection{Hole-doped cuprates}

Assuming that $\rho_{zz}$ scales with doping in the same way as the in-plane superfluid stiffness~\citep{Note2}, our results on the {\it hole-doped} side of the phase diagram in Figs.~\ref{fig:superfluid_stiffness_coexistence_U12_U8_tperp}~a) and \ref{fig:superfluid_stiffness_coexistence_U12_U8_tperp}~b) are consistent with the experimental drop of $\rho_{zz}(T=0)$ upon approaching half-filling in cuprates~\citep{Uemura_T_c_scaling_rho_s_several_layers_1989,Uemura_similarities_1991}. The quadratic component of the doping dependence that we found even seems consistent with the experimentally-inferred doping dependence of $\rho_{zz}(T=0)$~\cite{Hosseini_Broun_Sheehy_Davis_Franz_Hardy_Liang_Bonn_2004}. Earlier calculations of the in-plane superfluid stiffness that do not take into account antiferromagnetism explicitly have also found that superfluid stiffness vanishes near half filling when the Mott transition is taken into account. These calculations were done, for example, with slave-particles~\citep{LeeWen:1997,mallik2018surprises} for the $t-J$ model or with variational Monte Carlo for the Hubbard model~\citep{Paramekanti:2004}. 

The BCS-like drop in the superfluid stiffness that we find on the highly overdoped side is, however, not consistent with the linear doping dependence found experimentally in Refs.~\onlinecite{bozovic_overdoping_dependence_2016,Armitage_Stiffness:2019} if we assume that the in-plane superfluid stiffness measured in these experiments behaves in the same way as $\rho_{zz}$ calculated here. It has however been argued theoretically that the behavior of the superfluid stiffness on the overdoped side is consistent with BCS dirty $d$-wave behavior~\citep{Broun_dirty-d-wave:2017,Broun_Hirschfeld:2018}.

At intermediate values of $U$, for example $U=8t$ in Fig.~\ref{fig:superfluid_stiffness_coexistence_U12_U8_tperp}~b), antiferromagnetism plays an important role in making the superfluid stiffness vanish before half-filling. For larger clusters, it was found that superconductivity begins at a finite doping away from half-filling~\citep{Gull_Millis_Parcollet:2013}, even in the absence of antiferromagnetism. Nevertheless, comparing Fig.~\ref{fig:superfluid_stiffness_coexistence_U655_U5_tperp}~b) for $U$ below the critical $U_c$ for the Mott transition with Figs.~\ref{fig:superfluid_stiffness_coexistence_U12_U8_tperp}~a) and \ref{fig:superfluid_stiffness_coexistence_U12_U8_tperp}~b) for $U$ larger than $U_c$, it is clear that over most of the doping range the much smaller value of $\rho_{zz}$ and its doping dependence at large $U$ is controlled by Mott physics, not by competition with antiferromagnetism since antiferromagnetism appears only close to half-filling. 

Note however that our cluster can't accomodate long-period or incommensurate spin-density waves. These are seen both in experiments~\citep{Yamada_SDW:1998,Wakimoto_SDW:2000,Wakimoto_SDW:1999,Fujita_SDW:2002,Haug_2010} and in infinite-lattice calculations using methods that are valid for weak-~\citep{Yamase_Metzner_SDW:2016,Eberlein_Metzner_SDW:2016,Schultz_SDW_1990} to intermediate-strength interaction~\citep{Vilk:1997}. A preprint that appeared as this paper was prepared~\citep{Metzner_Stiffness:2019} obtains results similar to ours in the hole-doped regime using mean-field parameters obtained from functional renormalization group. Even though the superfluid stiffness is similar to ours, its fall towards half-filling is caused by coexistence with commensurate antiferromagnetism. Results in the incommensurate regime were not presented. For $U>U_c$ our results suggest, but do not prove, that it is possible for superfluid stiffness to control $T_c$ in the underdoped regime even when there is no coexisting antiferromagnetism, contrary to the results for weak interaction strength~\cite{Metzner_Stiffness:2019}.  

\subsection{Electron-doped cuprates}

It is in electron-doped cuprates that competition with antiferromagnetism is strongest and it is there also that coexistence occurs in our calculations. Even though electronic-structure calculations~\cite{weber_strength_2010} and comparisons of theory~\cite{senechal_hot_2004,kyung_pseudogap_2004} with photoemission~\cite{Armitage:2001,Armitage:2002} and neutron experiment~\cite{Motoyama2007_AFM_corr_more_prominent_e_doped} show that the value of $U$ should be in close vicinity to the Mott transition, this is not crucial for qualitative features of $\rho_{zz}$ as a function of doping. They are quite similar in the case of electron doping for $U=12$ in Fig.~\ref{fig:superfluid_stiffness_coexistence_U12_U8_tperp}~a), $U=8$ in Fig.~\ref{fig:superfluid_stiffness_coexistence_U12_U8_tperp}~b) and $U=6.55$ in Fig.~\ref{fig:superfluid_stiffness_coexistence_U655_U5_tperp}~a), which are all in the doped Mott insulator regime. In all cases: a) there is a small coexistence region where superfluid stiffness decreases rapidly compared with the value it would have in a pure superconducting state, b) antiferromagnetism overcomes completely superconductivity at a doping that is distinctly away from half-filling, as found in experiments~\cite{Luke_e-doped:1990,Greene_e-doped:2007, Motoyama2007_AFM_corr_more_prominent_e_doped,saadaoui_phase_2015} (see also note \footnote{When antiferromagnetism is destroyed in the annealing process~\cite{Horio_Protect_Anneal_e-doped_2016}, superconductivity ends very close to half-filling, which confirms that competition with antiferromagnetism is crucial in electron-doped superconductors}),
c) as one decreases doping, antiferromagnetism starts to coexist with superconductivity close to the doping where $\rho_{zz}$ reaches its maximum and d) the superfluid stiffness has a jump, or a rapid change in slope at smaller $U$, when one enters the coexistence phase from the pure superconducting phase at large doping. A similar jump was found in Ref.~\onlinecite{Metzner_Stiffness:2019}. 

One of the difficulties encountered by the one-band Hubbard or $t-J$ models has been that at zero temperature, when competition with long-range antiferromagnetic order is not allowed (pure regime), the size of the superconducting order parameter is larger on the electron-doped side of the phase diagram~\citep{Kancharla:2008,White_Scalapino_asymmetry:1999}. This is quite clear in three bottom subfigures of Figs.~\ref{fig:superfluid_stiffness_coexistence_U12_U8_tperp}~a) and \ref{fig:superfluid_stiffness_coexistence_U12_U8_tperp}~b). This was interpreted as a prediction that the transition temperature for electron-doped systems is larger than for hole-doped curates, contrary to observation. But one should not confuse the size of the superconducting order parameter with the value of the transition temperature. The correct value of the order parameters shows that competition with antiferromagnetic long-range order leads to the disappearance of superconductivity near half-filling, which decreases considerably the maximum value that the superconducting order-parameter can reach on the electron-doped side.  Assuming that the in-plane superfluid stiffness scales with doping in a way similar to the $c$-axis results, one notices that the superfluid stiffness at optimal doping, that can be dominant in determining the value of $T_c$, is in all cases smaller on the electron-doped than on the hole-doped side of the phase diagram. In addition, in the actual materials, the value of $U$ should be somewhat smaller for electron-doped cuprates, as mentioned above.

%%%%%%%%%%%%%%%%%%%%%%%%%%%%%%%%%%%%%%%%%%%%%%%% Conclusion %%%%%%%%%%%%%%%%%%%%%%%%%%%%%%%%%%%%%%%%%%%%%%%%%%%%%%%%%%%%%%%%%%
%%

\section{Conclusion}
\label{sec:conclusion}

We computed the $c$-axis superfluid stiffness at zero temperature for the one-band two-dimensional square-lattice Hubbard model. We solved the model on a $2\times 2$ plaquette using ED-CDMFT for model parameters appropriate for the cuprates. In finite-temperature $2\times 2$ plaquette calculations, the value of the superconducting transition temperature~\citep{Fratino2016} indicates the formation of local pairs, not necessarily the actual transition temperature, which, as our calculations suggest, is controlled by superfluid stiffness in the underdoped regime. 

On the hole-doped side, for YBCO band parameters and $U$ larger than the critical value for the Mott transition, it is mostly Mott physics that controls the value of the superfluid stiffness $\rho_{zz}$ near half-filling, although competition with antiferromagnetism does play a role just before half-filling. Superfluid stiffness along the $c$-axis increases with hole doping with linear plus quadratic dependence on doping, in qualitative agreement with experiment~\citep{Hosseini_Broun_Sheehy_Davis_Franz_Hardy_Liang_Bonn_2004}.

On the electron-doped side, our results suggest that it is the competition between AF and $d$SC that is most important even near optimal doping. This is suggested both by the value of the superconducting order parameter and by the superfluid stiffness $\rho_{zz}$ that jumps down~\cite{Metzner_Stiffness:2019} and then drops precipitously as soon as antiferromagnetism starts to coexist with superconductivity, a prediction for experiment. Just above that doping, $\rho_{zz}$ takes its largest value. This drop in $\rho_{zz}$ strongly depends on the electronic structure and on the value of the interaction $U$. The drop in $\rho_{zz}$ is more prominent for $U=8t$ and YBCO-like parameters. The reduction of $\rho_{zz}$ in the underdoped regime would increase the phase fluctuations of the superconducting order parameter. Hence, phase competition could be, according to the Uemura scaling relation~\citep{Uemura_T_c_scaling_rho_s_several_layers_1989}, an important factor in the depletion of $T_c$ in the underdoped regime for electron-doped cuprates as well. The disappearance of superconductivity closer to half-filling, however, comes from the fact that antiferromagnetism wins the competition with superconductivity in electron-doped cuprates.   

For both hole- and electron-doping at large $U$, the superfluid stiffness jumps extremely quickly to zero when the system becomes normal in the overdoped regime, in qualitative agreement with the expected BCS behavior. 

The effect of the in-plane modulation of the hopping amplitude along  the $c$-axis is important: in the electron-doped case, at large $U$ on the electron-doped side it increases $\rho_{zz}$ in the regime where only superconductivity exists while it decreases it when there is coexistence with antiferromagnetism. This is understood in terms of where the $d$-wave superconducting gap is important compared with the underlying state. 

We expect that competition with other types of order could have an effect on $\rho_{zz}$ similar to competition with antiferromagnetism. In future work, we plan to perform finite-temperature calculations to understand some of the unusual features of the superfluid-stiffness~\citep{Hosseini_Broun_Sheehy_Davis_Franz_Hardy_Liang_Bonn_2004} and its more precise role in determining the transition temperature. Even though it has a phase diagram very similar to the one-band model~\cite{FratinoChargeTransfer:2016}, we also plan to study the charge-transfer three-band model.

\begin{acknowledgments}
    We are grateful to Steve Kivelson, Giovanni Sordi and Simon Verret for discussions and to an anonymous referee for detailed constructive criticism. This work has been supported by the Natural Sciences and Engineering Research Council of Canada (NSERC) under grant RGPIN-2014-04584, the Canada First Research Excellence Fund and by the Research Chair in the Theory of Quantum Materials. Simulations were performed on computers provided by the Canadian Foundation for Innovation, the Minist\`ere de l'\'Education des Loisirs et du Sport (Qu\'ebec), Calcul Qu\'ebec, and Compute Canada.
\end{acknowledgments}

%--------------------------------------------------------------------------------------------------

\appendix

%--------------------------------------------------------------------------------------------------
\section{Derivation of $\rho_{ab}^{AF+SC}$}
\label{sec:appendice:Derivation_rho_AF+SC}

In this Appendix, we give further details on the calculation of the general superfluid stiffness $\rho_{zz}$ \eqref{eq:superfluid_stiffness:general_formulae:superfluid_stiffness_expanded} for the CDMFT calculation and for the state with AF+dSC coexistence Eq.~\eqref{eq:superfluid_stiffness:d_wave_superconductivity_AFM:superluid_stiffness_periodization}. We also explain the expression for the vertices \eqref{eq:vertices+} and how they are calculated when the perpendicular hopping amplitude depends on in-plane wave vectors.

% In this section, we demonstrate that the general superfluid stiffness $\bm{\rho}_{ab}$, defined as 

% \begin{align}
% \label{eq:cluster_green_function_periodization:gamma_ab_def_free_energy}
% \bm{\rho}_{ab}=-\frac{\beta}{V}\frac{\delta^2\mathcal{F}}{\delta A_a \delta A_b}\bigg\rvert_{\bm{A}=\bm{0}},
% \end{align} leads naturally to Eq.\eqref{eq:superfluid_stiffness:d_wave_superconductivity_AFM:superluid_stiffness_periodization} used in the AF+$d$SC coexistence regime. In Eq.\eqref{eq:cluster_green_function_periodization:gamma_ab_def_free_energy} $\mathcal{F}$ represents the free energy calculated with the dressed interacting Green's function in the plane, neglecting the Luttinger-Ward contribution
% %%%%%%%%%%
% \begin{align}
% \label{eq:cluster_green_function_periodization:non_interacting_free_energy}
% \mathcal{F} = -\text{Tr}\ln\left(-\mathcal{G}^{-1}(\tilde{k})\right),
% \end{align}
% %%%%%%%%% 
% and $a$ and $b$ run over the cartesian orthogonal axes $x,y,z$. $\text{Tr}\left[\cdots\right]$ expands as $\frac{1}{\beta N}\sum_{\tilde{k}}\sum_{\sigma}\text{tr}\left[\cdots\right]$, with $\text{tr}\left[\cdots\right]$ acting on the Nambu space and $N$ the number of unit cells. 

\subsection{General superfluid stiffness for CDMFT cluster}

To understand how the tensors $T_{lmn}$ occur in our calculations, it is simpler to start from a mean-field-like Hamiltonian where hybridization functions are replaced by order parameters. 

We start from the following lattice Green's function 
%%%%
\begin{align}
\label{eq:appendice:electronic_greens_function}
\mathcal{G}(\tilde{k})=\frac{1}{i\omega_n + \mu -\mathcal{H}^0_{\tilde{\bm{k}}}-\Sigma_c(i\omega_n)},
\end{align}
%%%%%%    
where the 4-vector is defined by $\tilde{k}\equiv (\tilde{\bm{k}},i\omega_n)$, $\mathcal{H}^0_{\tilde{\bm{k}}}$ is the mean-field-like Hamiltonian, i.e~quadratic in field operators, and $\Sigma_c$ is the cluster self-energy. 

\begin{figure}[h!]
    \begin{center}
    \includegraphics[clip=true,trim=0cm 0cm 0cm 0cm, width=0.6\columnwidth, height=0.6\columnwidth]{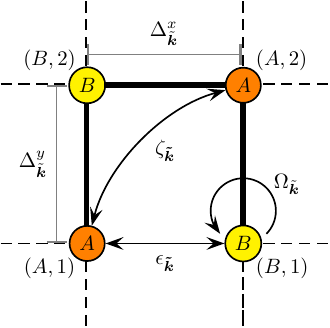}
    \caption{Schematic representation of the $2\times 2$ cluster cut out of the original lattice expressed in the orbital basis. The labels $A$ and $B$ on the sites account for the two sublattices resulting from AF order. We also illustrate the Fourier transforms of the of the nearest-neighbor $\epsilon_{\bm{\tilde{k}}}$, second nearest-neighbor $\zeta_{\bm{\tilde{k}}}$ and third nearest-neighbor hoppings $\Omega_{\bm{\tilde{k}}}$. These Fourier transforms take the same form as for the infinite, translationally invariant lattice. For clarity, there is no repetition of the various hoppings on the figure.}
    \label{fig:superfluid_stiffness:orbital_basis_spin_up}	
    \end{center}
    \end{figure}

In the CDMFT calculations, we consider a cluster consisting of 4 sites, therefore $N_c = 4$ in the expression for the spinor that we use as a basis (Eq.~\eqref{eq:Models_and_methods:ED_with_parametrization:cluster_site_basis_Green_function}):
 %%%%%%%%%%
 \begin{align}
    \label{eq:appendix:superfluid_stiffness_in_regime_of_coexistence_using_orbital_basis:general_Ham_T_basis}
    \hat{\Psi}_{\tilde{\bm{k}}}^{\dagger}=\begin{pmatrix}
    \hat{c}^{\dagger}_{\bm{\tilde{k}},\uparrow,A,1}&\hat{c}^{\dagger}_{\bm{\tilde{k}},\uparrow,A,2}&\hdots&\hat{c}_{-\bm{\tilde{k}},\downarrow,B,1}&\hat{c}_{-\bm{\tilde{k}},\downarrow,B,2}\\
    \end{pmatrix}.
    \end{align} 
%%%%%%%%%%%
Using the definitions in Fig.~\ref{fig:superfluid_stiffness:orbital_basis_spin_up}, the {\it mean-field} Nambu Hamiltonian then would be 
%%%%%%%%%%%%%%%%
 \begin{widetext}
    \begin{align}
        \label{eq:ch_rigidite_superfluide:formule_tracant_sur_amas:mixed_basis_Nambu_spinor_matrix_hamiltonian}
        &\mathcal{H}^{\alpha\beta;ab;rs}_{{MF}}(\tilde{\mathbf{k}}) =\notag\\
        &\begin{mpmatrix}
        \Omega_{\tilde{\mathbf{k}}}-M&\zeta_{\tilde{\mathbf{k}}}&\epsilon_{\tilde{\mathbf{k}}}&\epsilon_{\tilde{\mathbf{k}}}&\Delta_s&\Delta_{p,\tilde{\mathbf{k}}}&\Delta^x_{d,\tilde{\mathbf{k}}}&\Delta^y_{d,\tilde{\mathbf{k}}}\\
        \zeta_{\tilde{\mathbf{k}}}&\Omega_{\tilde{\mathbf{k}}}-M&\epsilon_{\tilde{\mathbf{k}}}&\epsilon_{\tilde{\mathbf{k}}}&\Delta_{p,\tilde{\mathbf{k}}}&\Delta_s&\Delta^y_{d,\tilde{\mathbf{k}}}&\Delta^x_{d,\tilde{\mathbf{k}}}\\
        \epsilon_{\tilde{\mathbf{k}}}&\epsilon_{\tilde{\mathbf{k}}}&\Omega_{\tilde{\mathbf{k}}}+M&\zeta_{\tilde{\mathbf{k}}}&\Delta^x_{d,\tilde{\mathbf{k}}}&\Delta^y_{d,\tilde{\mathbf{k}}}&\Delta_s&\Delta_{p,\tilde{\mathbf{k}}}\\
        \epsilon_{\tilde{\mathbf{k}}}&\epsilon_{\tilde{\mathbf{k}}}&\zeta_{\tilde{\mathbf{k}}}&\Omega_{\tilde{\mathbf{k}}}+M&\Delta^y_{d,\tilde{\mathbf{k}}}&\Delta^x_{d,\tilde{\mathbf{k}}}&\Delta_{p,\tilde{\mathbf{k}}}&\Delta_s\\
        \Delta^{\ast}_s&\Delta^{\ast}_{p,\tilde{\mathbf{k}}}&{\Delta^x}^{\ast}_{d,\tilde{\mathbf{k}}}&{\Delta^y}^{\ast}_{d,\tilde{\mathbf{k}}}&-\Omega_{-\tilde{\mathbf{k}}}+M&-\zeta_{-\tilde{\mathbf{k}}}&-\epsilon_{-\tilde{\mathbf{k}}}&-\epsilon_{-\tilde{\mathbf{k}}}\\
        \Delta^{\ast}_{p,\tilde{\mathbf{k}}}&\Delta^{\ast}_s&{\Delta^y}^{\ast}_{d,\tilde{\mathbf{k}}}&{\Delta^x}^{\ast}_{d,\tilde{\mathbf{k}}}&-\zeta_{-\tilde{\mathbf{k}}}&-\Omega_{-\tilde{\mathbf{k}}}+M&-\epsilon_{-\tilde{\mathbf{k}}}&-\epsilon_{-\tilde{\mathbf{k}}}\\
        {\Delta^x}^{\ast}_{d,\tilde{\mathbf{k}}}&{\Delta^y}^{\ast}_{d,\tilde{\mathbf{k}}}&\Delta^{\ast}_s&\Delta^{\ast}_{p,\tilde{\mathbf{k}}}&-\epsilon_{-\tilde{\mathbf{k}}}&-\epsilon_{-\tilde{\mathbf{k}}}&-\Omega_{-\tilde{\mathbf{k}}}-M&-\zeta_{-\tilde{\mathbf{k}}}\\
        {\Delta^y}^{\ast}_{d,\tilde{\mathbf{k}}}&{\Delta^x}^{\ast}_{d,\tilde{\mathbf{k}}}&\Delta^{\ast}_{p,\tilde{\mathbf{k}}}&\Delta^{\ast}_s&-\epsilon_{-\tilde{\mathbf{k}}}&-\epsilon_{-\tilde{\mathbf{k}}}&-\zeta_{-\tilde{\mathbf{k}}}&-\Omega_{-\tilde{\mathbf{k}}}-M
        \end{mpmatrix}.
        \end{align}
        
\end{widetext}
%%%%%%%%%%%
The superscripts in $\mathcal{H}^{\alpha\beta;ab;rs}_{{MF}}(\tilde{\mathbf{k}})$ take their meaning when the Hamiltonian is written as follows, taking advantage of the tensor-product form of the states on which the creation-annihilation operators \eqref{eq:appendix:superfluid_stiffness_in_regime_of_coexistence_using_orbital_basis:general_Ham_T_basis} act:  
\begin{align}
\label{eq:appendix:superfluid_stiffness_in_regime_of_coexistence_using_orbital_basis:general_Ham_T}
\bm{\mathcal{\hat{H}}}_{MF} &= \sum_{\bm{\tilde{k}}}\bigg(\sum_{l,m,n}A_{l,m,n}\sigma^l_{\alpha\beta}\tau_{ab}^m\tilde{\sigma}_{rs}^n\hat{c}^{\dagger}_{\bm{\tilde{k}},\alpha,a,r}\hat{c}_{\bm{\tilde{k}},\beta,b,s}\notag\\
&+\sum_{l^{\prime},m^{\prime},n^{\prime}}B_{l^{\prime},m^{\prime},n^{\prime}}\sigma_{\alpha\beta}^{l^{\prime}}\tau^{m^{\prime}}_{ab}\tilde{\sigma}_{rs}^{n^{\prime}}\hat{c}^{\dagger}_{\bm{\tilde{k}},\alpha,a,r}\hat{c}^{\dagger}_{\bm{\tilde{k}},\beta,b,s}+\text{H.c.}\bigg),
\end{align} 
%%%%%%%%%%%
where $\sigma^l_{\alpha\beta}$, $\tau_{ab}^m$ and $\tilde{\sigma}_{rs}^n$ are Pauli and identity matrices and $A$ and $B$ are order parameter tensors when mean-field is used. This structure of the Hamiltonian allowed us to introduce for short-hand in \autoref{subsec:superfluid_stiffness:general_formulae} the tensor   
%%%%%%%%%%%%%
\begin{align}
\label{eq:superfluid_stiffness:d_wave_superconductivity_AFM:T_matrices_representation_8d_spinor}
T_{lmn}=\sigma^l\otimes\tau^m\otimes\tilde{\sigma}^n.
\end{align} 
%%%%%%%%%%
Equation~\eqref{eq:appendix:superfluid_stiffness_in_regime_of_coexistence_using_orbital_basis:general_Ham_T}  represents the Hamiltonian before periodization to the AF-BZ. This is why there is a Pauli matrix $\tilde{\sigma}$. It is always diagonal in our case. 

We stress that we do not do mean-field theory. The effects of long-range order are all contained in the self-energy and hybridization function, not in the cluster Hamiltonian. 

Following the linear response procedure in Eq.~\eqref{eq:superfluid_stiffness:general_formulae:current_current_correlation_function} using the Green's function ~\eqref{eq:appendice:electronic_greens_function}, the formula obtained for the superfluid stiffness is 
%%%%%%%%%
\begin{align}
\label{eq:appendice:general_Gamma_ab_1st_expression}
\bm{\rho}_{ab} &= \frac{e^2}{\hbar^2\beta VN}\sum_{\tilde{\bm{k}},i\omega_n}\biggl(\text{tr}\left[\mathcal{G}(\tilde{k})\lambda^b_{\tilde{\bm{k}}}T_{300}\mathcal{G}(\tilde{k})\lambda^a_{\tilde{\bm{k}}}T_{300}\right]\notag\\
&+\text{tr}\left[\mathcal{G}(\tilde{k})\lambda^{ab}_{\tilde{\bm{k}}}\right]\biggr).
\end{align}
%%%%%%%%%% 
The derivation will become clearer below when we consider the AF+dSC mean-field state. 

The current vertices Eq.~\eqref{eq:vertices+} are:
%%%%
\begin{align}
    \lambda_{\bar{\bm{k}}}^i &\equiv \partial_{\bar{k}_i}\mathcal{H}_{\bar{\bm{k}},\sigma}^0\\
    \label{eq:vertices++}
    \lambda_{\bar{\bm{k}}}^{ji}&\equiv \partial_{\bar{k}_j}\partial_{\bar{k}_i}\mathcal{H}_{\bar{\bm{k}},\sigma}^0.
\end{align}
They can be obtained from the gradient of the kinetic-energy part of the Hamiltonian because the phase of the Fourier transform within a unit cell was taken into account when writing the Hamiltonian Eq.~\eqref{eq:ch_rigidite_superfluide:formule_tracant_sur_amas:mixed_basis_Nambu_spinor_matrix_hamiltonian} in the orbital basis~\cite{nourafkan_hall_2018}. The kinetic-energy part of the Hamiltonian, $\mathcal{H}_{\bar{\bm{k}},\sigma}^0$, is in the two 4$\times$4 diagonal blocks of Eq.~\eqref{eq:ch_rigidite_superfluide:formule_tracant_sur_amas:mixed_basis_Nambu_spinor_matrix_hamiltonian}. Because of the particle-hole transformation of down spins in the Nambu representation, we had to introduce a sign change through $\delta_{A_i} T_{000}= -\frac{e}{\hbar}\delta_{\bar{k}_i}T_{300}$ in Eq.~\eqref{eq:appendice:general_Gamma_ab_1st_expression} for the superfluid stiffness. In the CDMFT calculations, all off-diagonal terms are contained in the self-energy. 

%%%%%%%%%%%
%%%%%%%%%%%

% To obtain Eq.\eqref{eq:appendice:general_Gamma_ab_1st_expression}, the following relation was used: 
\subsection{Superfluid stiffness in the AF+dSC coexistence state}
 
Once again, the Pauli matrices $\sigma^l$ span the spin (Nambu) space, $\tau^m$ the AF sublattice space and $\tilde{\sigma}$ the sublattice spin projection degrees of freedom. In the infinite lattice with coexisting AF+dSC, the subspace spanned by $\tilde{\sigma}$ is not relevant. When we periodize the cluster Green's function onto the AF-BZ the superfluous $\tilde{\sigma}$ subspace disappears.

Let us go in more details through the derivation of $\rho_{zz}$ for a mean-field with AF+dSC microscopic coexistence. It will be clear how to handle the case of the periodized CDMFT Green's function Eq.~\eqref{eq:Models_and_methods:ED_with_parametrization:lattice_Green_function} mapped onto the AF-BZ (Fig.~\ref{fig:original_BZ_periodization}). 

The mean-field AF+dSC Hamiltonian Eq.~\eqref{eq:appendix:superfluid_stiffness_in_regime_of_coexistence_using_orbital_basis:general_Ham_T} would read, with $\{i,j\}=\{A,B\}$ and $\{\alpha,\beta\} = \{\uparrow,\downarrow\}$:

\begin{align}
\label{eq:appendix:superfluid_stiffness_in_regime_of_coexistence_using_orbital_basis:periodized_RBZ_Ham_T_basis}
\bm{\mathcal{\hat{H}}}_{MF}^{AF+dSC} &= -\sum_{ab}t_{ab}\sigma_{\alpha\beta}^0\tau_{ab}^1\hat{c}^{\dagger}_{\bm{k},\alpha,a}\hat{c}_{\bm{k},\beta,b}\notag\\
&+ M\sum_{ab}e^{i\bm{Q}\cdot\bm{r}_a+\phi}\sigma_{\alpha\beta}^3\tau_{ab}^3\hat{c}^{\dagger}_{\bm{k},\alpha,a}\hat{c}_{\bm{k},\beta,b}\notag\\
&+\sum_{ab}\Delta_{ab}\sigma_{\alpha\beta}^1\tau_{ab}^1\hat{c}^{\dagger}_{\bm{k},\alpha,a}\hat{c}^{\dagger}_{\bm{k},\beta,b} + \text{H.c.},
\end{align} where $t_{ab}$ is the hopping matrix between different AF sublattices, $\bm{Q}=(\pi,\pi)$ is the AF nesting wavevector and $\Delta_{ab} = \Delta$ if $\bm{r}_a-\bm{r}_b = \pm \bm{e}_x$, and $\Delta_{ab} = -\Delta$ if $\bm{r}_a-\bm{r}_b = \pm \bm{e}_y$, corresponding to $d_{x^2-y^2}$ pairing. The momentum vector $\bm{k}$ is defined in the rBZ.
%
%If one completes the Fourier transform into original Brillouin zone, one gets
%
%\begin{widetext}
%\begin{align}
%\label{eq:appendix:superfluid_stiffness_in_regime_of_coexistence_using_orbital_basis:original_Brillouin_Zone_H}
%\bm{\mathcal{\hat{H}}}^0_{\bm{k}} &= \sum_{\bm{k},\sigma}\epsilon_{\bm{k}}c_{\bm{k},\sigma}^{\dagger}c_{\bm{k},\sigma} +M\sum_{\bm{k}}\left[\left(c_{\bm{k},\uparrow}^{\dagger}c_{\bm{k}+\bm{Q},\uparrow}-c_{\bm{k},\downarrow}^{\dagger}c_{\bm{k}+\bm{Q},\downarrow}\right)+\text{H.c.} \right] +\sum_{\bm{k}}\bigg(\Delta_{\bm{k}} c_{-\bm{k},\downarrow}c_{\bm{k},\uparrow} +\Delta_{\bm{k}}^{\ast}c_{\bm{k},\uparrow}^{\dagger}c_{-\bm{k},\downarrow}^{\dagger}\bigg).
%\end{align}
%\end{widetext}

In orbital Nambu basis, the matrix form of the mean-field hamiltonian Eq.~\eqref{eq:appendix:superfluid_stiffness_in_regime_of_coexistence_using_orbital_basis:periodized_RBZ_Ham_T_basis} is
%\begin{align}
%\label{eq:appendix:superfluid_stiffness_in_regime_of_coexistence_using_orbital_basis:hamiltonian_orbital_basis_matrix_form}
%    \bm{\mathcal{\hat{H}}}^0_{\bm{k}} &=\notag\\
%    &\Scale[0.8]{\begin{blockarray}{ccccc}
%        A_{\uparrow} & B_{\uparrow} & A_{\downarrow} & B_{\downarrow} & \\ 
%    \begin{block}{[cccc]c}
%        \Omega_{\bm{k}}+\zeta_{\bm{k}}-M & \epsilon_{\bm{k}} & 0 & \Delta_{\bm{k}} & A_{\uparrow} \\
%        \epsilon_{\bm{k}} & \Omega_{\bm{k}}+\zeta_{\bm{k}}+M & \Delta_{\bm{k}} & 0 & B_{\uparrow} \\
%        0 & \Delta_{\bm{k}} & -\Omega_{-\bm{k}}-\zeta_{-\bm{k}}+M & -\epsilon_{-\bm{k}} & A_{\downarrow}  \\
%        \Delta_{\bm{k}} & 0 & -\epsilon_{-\bm{k}} & -\Omega_{\bm{k}}-\zeta_{-\bm{k}}-M & B_{\downarrow} \\
%    \end{block}
%    \end{blockarray}}
%\end{align} 
such that it can be written in terms of the SU$(2)\otimes$SU$(2)$ matrices $T_{lm}$ defined in Eq.~\eqref{eq:superfluid_stiffness:d_wave_superconductivity_AFM:T_matrices_representation}: 
%%%%%%%%%%%%
\begin{equation}
\label{eq:appendix:superfluid_stiffness_in_regime_of_coexistence_using_orbital_basis:mean_field_H_w_r_T}
\bm{\mathcal{H}}^{AF+dSC}_{MF}(\bm{k}) = \underbrace{(\zeta_{\bm{k}}+\Omega_{\bm{k}})}_{\xi_{\bm{k}}}T_{30}+\epsilon_{\bm{k}}T_{31}+\Delta_{\bm{k}} T_{11}-MT_{33}.
\end{equation} 
%%%%%%%%%%%%%

For the current $\langle j_a\rangle = -\frac{1}{V}\frac{\delta\mathcal{F}}{\delta A_a}$, one first needs
%%%%%%%%%%%%%%%%
\begin{align}
    \label{eq:appendix:superfluid_stiffness_in_regime_of_coexistence_using_orbital_basis:notation_superfluid_stiffness_formula_proof}
    \begin{split}
        \frac{\delta }{\delta A_a}T_{00}(\xi_{\bm{\bar{k}}}T_{30}+\epsilon_{\bm{\bar{k}}}T_{31})&=-\frac{e}{\hbar}\frac{\delta }{\delta \bar{k}_a}T_{30}(\xi_{\bm{\bar{k}}}T_{30}+\epsilon_{\bm{\bar{k}}}T_{31})
    \\
    &=-\frac{e}{\hbar}\frac{\delta }{\delta \bar{k}_a}(\xi_{\bm{\bar{k}}}T_{00}+\epsilon_{\bm{\bar{k}}}T_{01}).
    % \\
    % \frac{\delta \xi_{\bm{k}-\frac{e}{\hbar}\bm{A}T_{30}}}{\delta A_a}T_{30}&=-\frac{e}{\hbar}\frac{\delta \xi_{\bar{\bm{k}}}}{\delta \bar{k}_a}T_{00}
    % \\
    % \frac{\delta \epsilon_{\bm{k}-\frac{e}{\hbar}\bm{A}T_{30}}}{\delta A_a}T_{31}&=-\frac{e}{\hbar}\frac{\delta \epsilon_{\bar{\bm{k}}}}{\delta \bar{k}_a}T_{01}.
    \end{split}
    \end{align} 
%%%%%%%%%%%%%%%%%
The bar over $\bm{k}$ reminds us that the vector potential is contained in the wave-vector with a sign that differs between up and down spins. The current then can be written as  
%%%%%%%%%%%%%%%%
%\begin{widetext}
\begin{align}
\label{eq:appendix:superfluid_stiffness_in_regime_of_coexistence_using_orbital_basis:mean_current}
\langle j_a\rangle
% &=\frac{1}{\beta V}\sum_{k}\text{tr}\left[\bm{\mathcal{G}}(\bar{k})\left(\frac{\delta \xi_{\bm{k}-\frac{e}{\hbar}\bm{A}T_{30}}}{\delta A_a}T_{30}+\frac{\delta \epsilon_{\bm{k}-\frac{e}{\hbar}\bm{A}T_{30}}}{\delta A_a}T_{31}\right)\right]\notag \\ 
&=\frac{e}{\hbar\beta V}\sum_{\bar{k}}\text{tr}\left[\bm{\mathcal{G}}(\bar{k})\left(\frac{\delta \xi_{\bar{\bm{k}}}}{\delta \bar{k}_a}T_{00}+\frac{\delta \epsilon_{\bar{\bm{k}}}}{\delta \bar{k}_a}T_{01}\right)\right],
\end{align}
%\end{widetext} 
%%%%%%%%%%%%%%%
where one can use either the mean-field or the periodized CDMFT Green's function and where $\text{tr}[\cdots]$ operates in the 4$\times$4 Nambu space. We have supposed that the system is invariant under inversion ($\bm{k} = -\bm{k}$). 
%%%%%%%%%%%%%%%%
% \begin{align}
% \label{eq:appendix:superfluid_stiffness_in_regime_of_coexistence_using_orbital_basis:notation_superfluid_stiffness_formula_proof}
% \begin{split}
%     \frac{\delta }{\delta A_a}T_{00}(\xi_{\bm{k}}T_{30}+\epsilon_{\bm{k}}T_{31})&=-\frac{e}{\hbar}\frac{\delta }{\delta \bar{k}_a}T_{30}(\xi_{\bm{k}}T_{30}+\epsilon_{\bm{k}}T_{31})
% \\
% &=-\frac{e}{\hbar}\frac{\delta }{\delta \bar{k}_a}(\xi_{\bm{k}}T_{00}+\epsilon_{\bm{k}}T_{01})
% \\
% \frac{\delta \xi_{\bm{k}-\frac{e}{\hbar}\bm{A}T_{30}}}{\delta A_a}T_{30}&=-\frac{e}{\hbar}\frac{\delta \xi_{\bar{\bm{k}}}}{\delta \bar{k}_a}T_{00}
% \\
% \frac{\delta \epsilon_{\bm{k}-\frac{e}{\hbar}\bm{A}T_{30}}}{\delta A_a}T_{31}&=-\frac{e}{\hbar}\frac{\delta \epsilon_{\bar{\bm{k}}}}{\delta \bar{k}_a}T_{01}.
% \end{split}
% \end{align} 
%%%%%%%%%%%%%%%%%

The periodized CDMFT Green's function takes the same form as in Eq.~\ref{eq:appendice:electronic_greens_function} except that  
%%%%%%%%%%%%%
% \begin{align}
% \label{eq:appendix:superfluid_stiffness_in_regime_of_coexistence_using_orbital_basis:general_electronic_greens_function}
% \bm{\mathcal{G}}(k)=\frac{1}{i\omega_n+\mu-\bm{\mathcal{H}}^0_{\bm{k}}-\bm{\Sigma}_c(i\omega_n)},
% \end{align} 
%%%%%%%%%%%%%
$\bm{\mathcal{H}}^0_{\bm{k}}$ depends on $\bm{k}$ instead of $\bm{\tilde{k}}$ and has a smaller size since it contains only the $T_{30}$ and $T_{31}$ parts of the mean-field Hamiltonian Eq.~\eqref{eq:appendix:superfluid_stiffness_in_regime_of_coexistence_using_orbital_basis:mean_field_H_w_r_T}. All off-diagonal pieces are in the self-consistent off-diagonal self-energies. 

Inserting either the mean-field or CDMFT periodized Green's function%(Eq.~\eqref{eq:appendix:superfluid_stiffness_in_regime_of_coexistence_using_orbital_basis:general_electronic_greens_function})
, neglecting vertex corrections (i.e. the self-energy dependence of the vector potential $\bm{A}$), and using
%%%%%%%%%
\begin{align}
\label{eq:appendice:partial_derivative_green_function_handy_relation}
\frac{\delta \mathcal{G}(\tilde{k})}{\delta A_b} = -\mathcal{G}(\tilde{k})\frac{\delta \mathcal{G}^{-1}(\tilde{k})}{\delta A_b}\mathcal{G}(\tilde{k})
\end{align} 
%%%%%%%%%%
one can compute $\bm{\rho}_{ab}=-\frac{\delta \langle j_a\rangle}{\delta A_b}\big\rvert_{\bm{A}=\bm{0}}$:
%%%%%%%%%%
\begin{widetext}
\begin{align}
\label{eq:appendix:superfluid_stiffness_in_regime_of_coexistence_using_orbital_basis:gamma_AFM_d-SC_expanded}
\bm{\rho}_{ab}=&\frac{e^2}{\hbar^2\beta VN}\sum_{\bar{k}}\text{tr}\left[\bm{\mathcal{G}}(\bar{k})\left(\frac{\delta^2\xi_{\bar{\bm{k}}}}{\delta \bar{k}_b\delta \bar{k}_a}T_{30}+\frac{\delta^2\epsilon_{\bar{\bm{k}}}}{\delta \bar{k}_b \delta \bar{k}_a}\underbrace{T_{30}T_{01}}_{T_{31}}\right)\right]\bigg\rvert_{\bm{A}=\bm{0}}+\notag \\
&\frac{e^2}{\hbar^2\beta VN}\sum_{\bar{k}}\text{tr}\left[\bm{\mathcal{G}}(\bar{k})\left(\frac{\delta \xi_{\bar{\bm{k}}}}{\delta \bar{k}_b}T_{00}+\frac{\delta \epsilon_{\bar{\bm{k}}}}{\delta \bar{k}_b}T_{01}\right)\bm{\mathcal{G}}(\bar{k})\left(\frac{\delta \xi_{\bar{\bm{k}}}}{\delta \bar{k}_a}T_{00}+\frac{\delta \epsilon_{\bar{\bm{k}}}}{\delta \bar{k}_a}T_{01}\right)\right]\bigg\rvert_{\bm{A}=\bm{0}}.
\end{align}
\end{widetext} 
% The bare current vertex functions were defined as $\partial_{\tilde{k}_i}\mathcal{G}^{-1}(\tilde{k})\equiv -\lambda^i_{\tilde{\bm{k}}}$ and $\partial_{\tilde{k}_j}\partial_{\tilde{k}_i}\mathcal{G}^{-1}(\tilde{k})\equiv -\lambda^{ji}_{\tilde{\bm{k}}}$.
%%%%%%%%%%
The second term, so-called paramagnetic term, was obtained from the derivative of the Green's function Eq.~\eqref{eq:appendice:partial_derivative_green_function_handy_relation}. Once the partial derivatives have acted, we set $\bm{A}\to \bm{0}$.

%%%%%%%%%%%
\begin{figure}[h!]
    \begin{center}
    \includegraphics[clip=true,trim=0cm 0cm 0cm 0cm, width=0.8\columnwidth, height=0.3\textwidth]{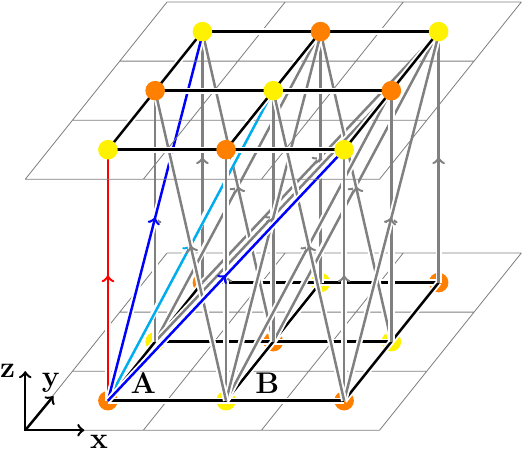}
    \caption{Example of stacked CuO$_2$ planes along the $c$-axis ($z$-axis). The different AF sublattices $A$ and $B$ are shown in orange and yellow, respectively (\textit{cf.}~Fig.~\ref{fig:superfluid_stiffness:orbital_basis_spin_up}). The red arrow illustrates a nearest-neighbor hopping, the cyan arrow a second neighbor hopping and the blue arrows third neighbor hoppings between two stacked CuO$_2$ planes. To lighten the figure, only half of the overall second- and third- neighbor hoppings are shown and a wide range of hoppings are coloured gray. All the hopping terms contained in $t_{\perp}$ (Eq.~\eqref{eq:superfluid_stiffness:general_formulae:bilayer_hopping_term}) shift electrons from one AF sublattice to another when hopping from one plane to another.}
    \label{fig:superfluid_stiffness:stacked_CuO2_planes_c_axis}	
    \end{center}
    \end{figure} 
    %%%%%%%%%%%

It is convenient to use the periodicity of the Brillouin zone to do a partial integration of the diamagnetic components 
%%%%%%
\begin{align}
    \label{eq:appendix:superfluid_stiffness_in_regime_of_coexistence_using_orbital_basis:gamma_AFM_d-SC_Brillouin_zone_periodicity_integral_parts}
    \int\frac{\mathrm{d}^2k}{\left(2\pi\right)^2}\frac{\partial^2 \xi_{\bm{k}}}{\partial k_b \partial k_a}&\text{tr}\left[\bm{\mathcal{G}}(k)T_{30}\right]\notag \\ = &-\int\frac{\mathrm{d}^2k}{\left(2\pi\right)^2}\frac{\partial \xi_{\bm{k}}}{\partial k_a}\text{tr}\left[\frac{\partial \bm{\mathcal{G}}(k)}{\partial k_b}T_{30}\right].
    \end{align} 
%%%%%%%%    
That allows a grouping of the diamagnetic and paramagnetic terms of Eq.~\eqref{eq:appendix:superfluid_stiffness_in_regime_of_coexistence_using_orbital_basis:gamma_AFM_d-SC_expanded},
Indeed, benefiting again from Eq.~\eqref{eq:appendice:partial_derivative_green_function_handy_relation} for the derivative of a Green's function (with the replacement $A_b\to k_b$), the final expression for the superfluid stiffness in the AF-$d$SC coexistence regime takes the form ($\bm{\mathcal{G}}(k)\to \bm{\mathcal{G}}$)
%%%%%%%%%%%
\begin{align}
\label{eq:appendix:superfluid_stiffness_in_regime_of_coexistence_using_orbital_basis:gamma_AFM_d-SC}
\bm{\rho}_{ab}&=\frac{e^2}{\hbar^2\beta VN}\times \notag\\
\sum_{k}&\Bigg[\frac{\partial \xi_{\bm{k}}}{\partial k_b}\frac{\partial \xi_{\bm{k}}}{\partial k_a}\bigg(\text{tr}\left[\bm{\mathcal{G}}T_{00}\bm{\mathcal{G}}T_{00}\right]-\text{tr}\left[\bm{\mathcal{G}}T_{30}\bm{\mathcal{G}}T_{30}\right]\bigg)\notag\\
&+\frac{\partial\xi_{\bm{k}}}{\partial k_b}\frac{\partial \epsilon_{\bm{k}}}{\partial k_a}\bigg(\text{tr}\left[\bm{\mathcal{G}}T_{00}\bm{\mathcal{G}}T_{01}\right]-\text{tr}\left[\bm{\mathcal{G}}T_{30}\bm{\mathcal{G}}T_{31}\right]\bigg)\notag \\
&+\frac{\partial \epsilon_{\bm{k}}}{\partial k_b}\frac{\partial \xi_{\bm{k}}}{\partial k_a}\bigg(\text{tr}\left[\bm{\mathcal{G}}T_{01}\bm{\mathcal{G}}T_{00}\right]-\text{tr}\left[\bm{\mathcal{G}}T_{31}\bm{\mathcal{G}}T_{30}\right]\bigg)\notag\\
&+\frac{\partial\epsilon_{\bm{k}}}{\partial k_b}\frac{\partial \epsilon_{\bm{k}}}{\partial k_a}\bigg(\text{tr}\left[\bm{\mathcal{G}}T_{01}\bm{\mathcal{G}}T_{01}\right]-\text{tr}\left[\bm{\mathcal{G}}T_{31}\bm{\mathcal{G}}T_{31}\right]\bigg)\Bigg].
\end{align} 
%%%%%%%%
Equation \eqref{eq:appendix:superfluid_stiffness_in_regime_of_coexistence_using_orbital_basis:gamma_AFM_d-SC} is general if vertex corrections are neglected and $a,b\in\{x,y,z\}$. The Green's functions obtained from periodizing the CDMFT solutions can be introduced where $\bm{\mathcal{G}}$ stands in Eq.~\eqref{eq:appendix:superfluid_stiffness_in_regime_of_coexistence_using_orbital_basis:gamma_AFM_d-SC}. If one does not allow for symmetry breaking associated with antiferromagnetism, one retrieves the superfluid stiffness formula for pure superconducting systems Eq.~\eqref{eq:superfluid_stiffness:superfluid_stiffness_SC_only}. 

\subsection{Vertices for $\rho_{zz}$ when there is a dependence on in-plane wave vectors}
%%%%%%%%%
Since we compute $\rho_{zz}$, we do not need all the terms of Eq.~\eqref{eq:appendix:superfluid_stiffness_in_regime_of_coexistence_using_orbital_basis:gamma_AFM_d-SC}. Fourier transforming the perpendicular hopping Eq.~\eqref{eq:superfluid_stiffness:general_formulae:bilayer_hopping_term} back to lattice coordinates, one can see that there are three different interlayer hopping terms involved in Eq.~\eqref{eq:superfluid_stiffness:general_formulae:bilayer_hopping_term} and they all make the electrons hop from one AF sublattice to the other, as can be seen from Fig.~\ref{fig:superfluid_stiffness:stacked_CuO2_planes_c_axis}. Hence, only the last term of Eq.~\eqref{eq:appendix:superfluid_stiffness_in_regime_of_coexistence_using_orbital_basis:gamma_AFM_d-SC} remains after setting $a=b=z$.

%\bibliography{Bibliography}


\end{document}

% Title layout
\title{Superfluid stiffness in cuprates: Effect of Mott transition and phase competition --- Supplementary Materials}
\author{O. \surname{Simard}}
%\email[Corresponding author: ]{simon.verret@usherbrooke.ca}
\affiliation{D\'epartement de physique and Institut quantique, Universit\'e de Sherbrooke, Qu\'ebec, Canada  J1K 2R1}
\author{C.-D. \surname{Hébert}}
\affiliation{D\'epartement de physique and Institut quantique, Universit\'e de Sherbrooke, Qu\'ebec, Canada  J1K 2R1}
\author{A. \surname{Foley}}
\affiliation{D\'epartement de physique and Institut quantique, Universit\'e de Sherbrooke, Qu\'ebec, Canada  J1K 2R1}
\author{D. \surname{S\'en\'echal}}
\affiliation{D\'epartement de physique and Institut quantique, Universit\'e de Sherbrooke, Qu\'ebec, Canada  J1K 2R1}
\author{A.-M. S. \surname{Tremblay}}
\affiliation{D\'epartement de physique and Institut quantique, Universit\'e de Sherbrooke, Qu\'ebec, Canada  J1K 2R1}
\affiliation{Canadian Institute for Advanced Research, Toronto, Ontario, Canada M5G 1Z8}
\date{\today}
\keywords{}
%  Abstract
\begin{abstract}
Here we present the results for the superfluid stiffness $\rho_{\parallel}$ computed in the CuO$_2$ plane using ED-CDMFT. The parameters $U$, $t^{\prime}$ and $t^{\prime\prime}$ are the same as those in the main text. We observe that, apart from an overall scale factor, there is a qualitative agreement between the $c$-axis superfluid stiffness $\rho_{zz}$ and the in-plane one $\rho_{\parallel}$. The interplay of AF and $d$SC decreases $\rho_{\parallel}$ as soon as AF order sets in. The calculations were performed neglecting the current vertex corrections. These are usually negligible in broken-symmetry states where a gap has opened, except when conservation laws are involved. When Kosterlitz-Thouless physics is important, $\rho_{\parallel}$ can be used to set a bound on $T_c$~\cite{Esterlis2018_Nature_Tc_bound}.  
\end{abstract}
\maketitle

%%%%%%%%%%%%%%%%%%%%%%%%%%%%%%%%%%%%%%%%%%%%%%%% Introduction %%%%%%%%%%%%%%%%%%%%%%%%%%%%%%%%%%%%%%%%%%%%%%%%%%%%%%%%%%%%%%%%%%
%%

\section{Introduction}
\label{sec:Introduction_Supp}

Cuprates are highly anisotropic layered materials: electronic correlations are strong within the CuO$_2$ planes, whilst the planes are weakly coupled along the axis perpendicular to their surface ($c$ axis). In the main part of this paper, we presented results of calculations for the $c$-axis superfluid stiffness. What about the in-plane superfluid stiffness, $\rho_{\parallel}$? To be rigourous and obtain quantitative agreement with experiment, one would have to compute vertex corrections. These corrections are computationally expensive and can usually be neglected in gapped broken symmetry states, unless conservation laws are relevant for the response~\citep{schrieffer2018theory}. 

We use the same method and model as in the main text. The major change occurs in the expression for the superfluid stiffness in the regime of AF+$d$SC coexistence. Due to the current vertices, Eq.(23) is no longer valid. In the following, we introduce the formula to compute $\rho_{\parallel}$ and explain its components.

%%%%%%%%%%%%%%%%%%%%%%%%%%%%%%%%%%%%%%%%%%%%%%%% Superfluid Stiffness %%%%%%%%%%%%%%%%%%%%%%%%%%%%%%%%%%%%%%%%%%%%%%%%%%%%%%%%%%%
%%

\begin{figure*}
    \includegraphics[clip=true,trim=0cm 0cm 0cm 0cm, width=\columnwidth ]{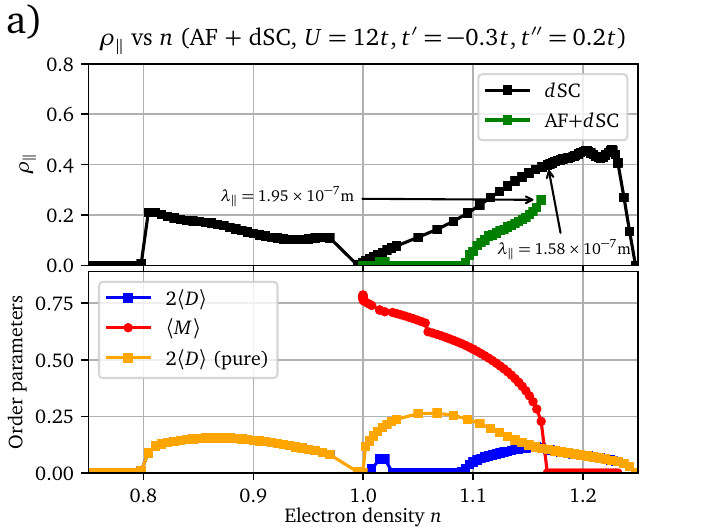}
    \includegraphics[clip=true,trim=0cm 0cm 0cm 0cm,width=\columnwidth ]{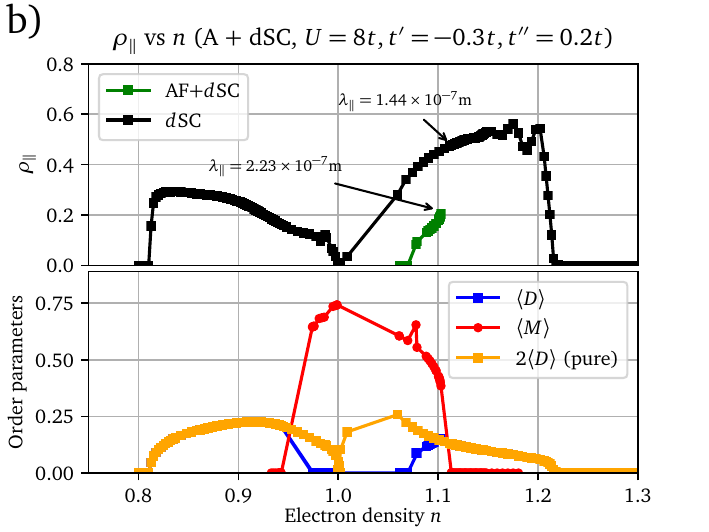}
    \caption{a) In-plane superfluid stiffness ($\rho_{\parallel}$) as a function of density ($n$) for $U = 12t$, $t^{\prime} = -0.3t$ and $t^{\prime\prime} = 0.2t$. The green markers show $\rho_{\parallel}$ in the coexistence regime along either the $x$-axis 
    or $y$-axis. The black markers illustrate $\rho_{\parallel}$ in the case of pure SC, along either the 
    $x$-axis or $y$-axis. Some values of the in-plane London penetration length ($\lambda_{\parallel}$) are 
    shown. The bottom subfigure illustrates the $d$SC order parameter in the pure regime $\langle D\rangle\ (\text{pure})$ (orange), and the $d$SC $\langle D\rangle$ (blue) and AF $\langle M\rangle$ (red) order parameters as a function of $n$ in the coexistence regime. b) $\rho_{\parallel}$ as a function of $n$ for $U=8t$, $t^{\prime}=-0.3t$ and $t^{\prime\prime}=0.2t$. The figure layout 
    is as Fig.~\ref{fig:superfluid_stiffness:Supp:in_plane_U12_U8_rho}~a).}
    \label{fig:superfluid_stiffness:Supp:in_plane_U12_U8_rho}	
\end{figure*}

\section{Superfluid Stiffness}
\label{sec:Superfluid_Stiffness_Supp}

The formula for $\rho_{\parallel}$ in the AF+$d$SC coexistence regime is essentially the same as Eq.~(A16), with $a,b\in\{x,y\}$

\begin{align}
\label{eq:appendix:superfluid_stiffness_in_regime_of_coexistence_using_orbital_basis:gamma_AFM_d-SC}
\bm{\rho}^{\parallel}_{ab}&=\frac{e^2}{\hbar^2\beta VN}\times \notag\\
\sum_{k}&\Bigg[\frac{\partial \xi_{\bm{k}}}{\partial k_b}\frac{\partial \xi_{\bm{k}}}{\partial k_a}\bigg(\text{tr}\left[\bm{\mathcal{G}}T_{00}\bm{\mathcal{G}}T_{00}\right]-\text{tr}\left[\bm{\mathcal{G}}T_{30}\bm{\mathcal{G}}T_{30}\right]\bigg)\notag\\
&+\frac{\partial\xi_{\bm{k}}}{\partial k_b}\frac{\partial \epsilon_{\bm{k}}}{\partial k_a}\bigg(\text{tr}\left[\bm{\mathcal{G}}T_{00}\bm{\mathcal{G}}T_{01}\right]-\text{tr}\left[\bm{\mathcal{G}}T_{30}\bm{\mathcal{G}}T_{31}\right]\bigg)\notag \\
&+\frac{\partial \epsilon_{\bm{k}}}{\partial k_b}\frac{\partial \xi_{\bm{k}}}{\partial k_a}\bigg(\text{tr}\left[\bm{\mathcal{G}}T_{01}\bm{\mathcal{G}}T_{00}\right]-\text{tr}\left[\bm{\mathcal{G}}T_{31}\bm{\mathcal{G}}T_{30}\right]\bigg)\notag\\
&+\frac{\partial\epsilon_{\bm{k}}}{\partial k_b}\frac{\partial \epsilon_{\bm{k}}}{\partial k_a}\bigg(\text{tr}\left[\bm{\mathcal{G}}T_{01}\bm{\mathcal{G}}T_{01}\right]-\text{tr}\left[\bm{\mathcal{G}}T_{31}\bm{\mathcal{G}}T_{31}\right]\bigg)\Bigg],
\end{align} where $\epsilon_{\mathbf{k}}$ represents the dispersion arising from nearest-neighbor hopping and $\xi_{\mathbf{k}}$ the dispersion arising from second and third nearest-neighbor hoppings (see Fig.~5). The trace ``$\text{tr}$'' is performed over the enlarged Nambu space described by Eq.~(9), and the tensors $T$ are introduced in Eq.~(22). The Green's functions $\bm{\mathcal{G}}$ implicitely depend on the spin and the 4-vector $k$, and their self-energy is calculated with the ED-CDMFT procedure described in Section II.B. The $\bm{\mathcal{G}}$'s are periodized in the rBZ according to Eq.~(11). Results for $\rho_{\parallel}$ using the cumulant periodization scheme are shown in Ref.~\onlinecite{simard_master}. Results for $\rho_{\parallel}$ are also shown in Ref.~\onlinecite{simard_master} when not periodizing at all.

An intuitive understanding of the origin of each term of Eq.~\eqref{eq:appendix:superfluid_stiffness_in_regime_of_coexistence_using_orbital_basis:gamma_AFM_d-SC} can be useful. It can be deduced from the vertices and from the indices of the $T$ matrices.  %Therefore, we break down Eq.~\eqref{eq:appendix:superfluid_stiffness_in_regime_of_coexistence_using_orbital_basis:gamma_AFM_d-SC} into its components. 
All the scattering processes they describe happen within one CuO$_2$ plane. The first term contains the contributions to $\rho_{\parallel}$ from the scattering of two Cooper pairs that conserve their  quantum numbers $A,B$. %and $\uparrow,\downarrow$. 
The second and third terms of Eq.~\eqref{eq:appendix:superfluid_stiffness_in_regime_of_coexistence_using_orbital_basis:gamma_AFM_d-SC} are equivalent, as they both account for the scattering processes of Cooper pairs that change the sublattice quantum number ($A,B$) of one Cooper pair compared with the initial state. Finally, the contribution coming from the last term is responsible for scattering processes that change the sublattice quantum number of both Cooper pairs. This fourth term was the only one contributing to $\rho_{zz}$, as can be seen from Eq.~(23), Fig.~6, and the discussion in Appendix A. This is because hopping between the planes occurs only between different sublattices.

A good way to assess the validity of Eq.~\eqref{eq:appendix:superfluid_stiffness_in_regime_of_coexistence_using_orbital_basis:gamma_AFM_d-SC} is to set the AF order parameter $\langle M\rangle$ to $0$. Doing so, one should retrieve the pure $d$SC expression Eq.~(20) with the full bare velocity. Keeping in mind that $\epsilon(\bm{k})$ 
%partial derivatives $\frac{\partial \epsilon_{\mathbf{k}}}{\partial k_x}$ and $\frac{\partial \epsilon_{\mathbf{k}}}{\partial k_y}$ 
changes sign when crossing the AF zone boundary, this is the case. For that reason, $\rho_{\parallel}$ for both the pure SC and coexisting states has been computed using Eq.~\eqref{eq:appendix:superfluid_stiffness_in_regime_of_coexistence_using_orbital_basis:gamma_AFM_d-SC}.

%%%%%%%%%%%%%%%%%%%%%%%%%%%%%%%%%%%%%%%%%%%%%%%% Results %%%%%%%%%%%%%%%%%%%%%%%%%%%%%%%%%%%%%%%%%%%%%%%%%%%%%%%%%%%%%%%%%%
%%

\section{Results}
\label{sec:Results_Supp}

\begin{figure*}
    \includegraphics[clip=true,trim=0cm 0cm 0cm 0cm, width=\columnwidth ]{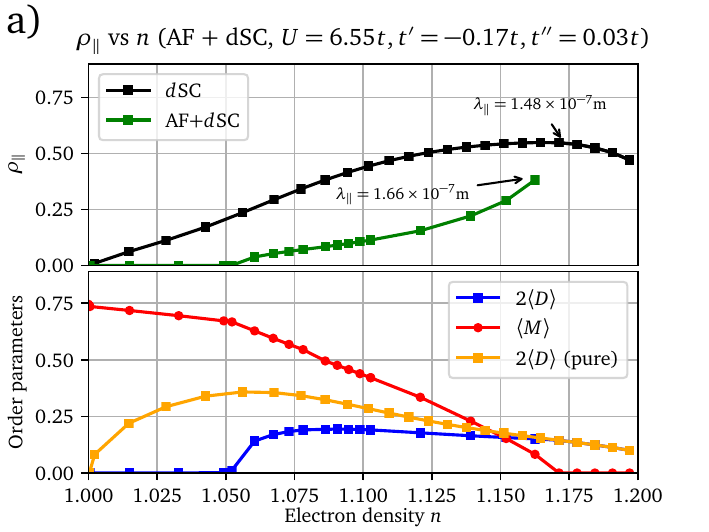}
    \includegraphics[clip=true,trim=0cm 0cm 0cm 0cm,width=\columnwidth ]{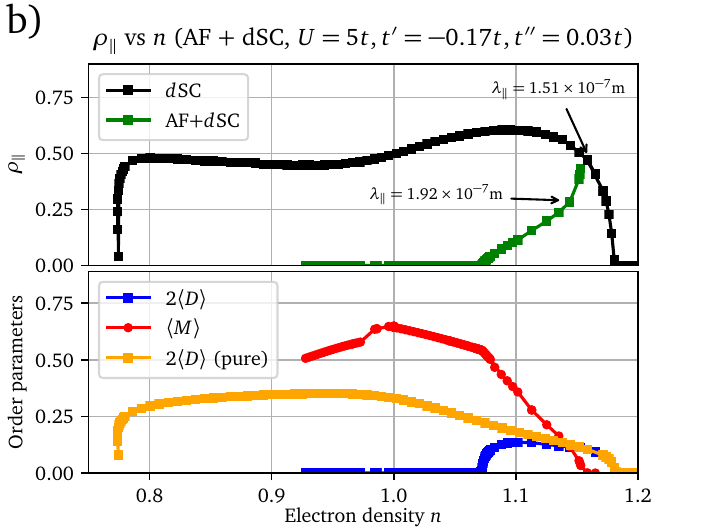}
    \caption{a) $\rho_{\parallel}$ as a function of $n$ for $U=6.55t$, $t^{\prime}=-0.17t$ and $t^{\prime\prime}=0.03t$. The figure 
    layout is as fig.~\ref{fig:superfluid_stiffness:Supp:in_plane_U12_U8_rho}. b) $\rho_{\parallel}$ as a function of $n$ for $U=5t$, $t^{\prime}=-0.17t$ and $t^{\prime\prime}=0.03t$. The figure 
    layout is as in fig.~\ref{fig:superfluid_stiffness:Supp:in_plane_U12_U8_rho}.}
    \label{fig:superfluid_stiffness:Supp:in_plane_U655_U5_rho}	
\end{figure*}

In this section, we show $\rho_{\parallel}$ for different sets of parameters: Figs.~\ref{fig:superfluid_stiffness:Supp:in_plane_U12_U8_rho}~a) and \ref{fig:superfluid_stiffness:Supp:in_plane_U12_U8_rho}~b) illustrate $\rho_{\parallel}$ for $U=12t>U_c$ and $U=8t>U_c$, for the YBCO-like set of band parameters. Figs.~ \ref{fig:superfluid_stiffness:Supp:in_plane_U655_U5_rho}~a) and \ref{fig:superfluid_stiffness:Supp:in_plane_U655_U5_rho}~b) illustrate $\rho_{\parallel}$ for $U=6.55t>U_c$ and $U=5t<U_c$, for the set of NCCO-like band parameters. The drop of $\rho_{\parallel}$ with the onset of $\langle M\rangle$ is a feature shared by all the cases. That drop is more pronounced for the YBCO-like parameters. For fixed set of parameters, all the values of the in-plane penetration depth $\lambda_{\parallel}$ that have been calculated are lower than $\lambda_{c}$, as expected. For these values of $\lambda_{\parallel}$ and $\lambda_{c}$ in physical units, we have used the numbers at the end of Section III.A. However, on the plots in the main paper and in the supplementary materials, perpendicular hopping $t_{bi}$ and in-plane nearest-neighbor hopping $t$ are both unity. This may mislead the reader to think that both superfluid stiffnesses are of the same order of magnitude.   

The results for $\rho_{\parallel}$ are more sensitive to translational symmetry breaking associated with the supercluster network. Periodization in the AF Brillouin zone gives differences between the two perpendicular directions that can be as large as 25\%. The results in the figures have been averaged between the two directions. Since the current vertices in Eq.~\eqref{eq:appendix:superfluid_stiffness_in_regime_of_coexistence_using_orbital_basis:gamma_AFM_d-SC} have a more complex $k$-space dependence, this may amplify the translational symmetry breaking caused by the periodization. %Using  values of $\rho_{\parallel}$.

%%%%%%%%%%%%%%%%%%%%%%%%%%%%%%%%%%%%%%%%%%%%%%%% Conclusion %%%%%%%%%%%%%%%%%%%%%%%%%%%%%%%%%%%%%%%%%%%%%%%%%%%%%%%%%%%%%%%%%%
%%

\section{Conclusion}
\label{sec:conclusion}

Similarly to the results for $\rho_{zz}$, the in-plane superfluid stiffness $\rho_{\parallel}$ drops precipitously once the AF order sets in. Sometimes, this happens in a first-order-like manner. The calculations along the three different axes of the unit-cell give the same qualitative dependence on doping at zero-temperature. In a subsequent publication, we are going to discuss bounds on the superconducting transition temperature that can be set using the results of these supplementary materials.

\pagebreak    

%\bibliography{Bibliography}
